\documentclass[
 showpacs,
 superscriptaddress,
 aps,
 pra,
 amsmath,amssymb,nofootinbib,
 reprint,
]{revtex4-1}

\usepackage[english]{babel}
\usepackage[T1]{fontenc}
\usepackage{hyperref}
\usepackage{graphicx}
\usepackage{subfigure}
\usepackage{amsmath}
\usepackage{amssymb}
\usepackage{slashed}
\usepackage{dsfont}

\newcommand{\dint}{\text{d}}
\newcommand{\Nc}{N_{\rm c}}
\newcommand{\Nf}{N_{\rm f}}
\newcommand{\Unit}{{\mathds 1}}
\newcommand{\psib}{\bar{\psi}}
\newcommand{\dslash}{\slashed{\partial}}
\newcommand{\Tr}{{\rm Tr}}
\renewcommand{\i}{{\rm i}}
\newcommand{\e}{{\rm e}}
\newcommand{\eps}{\varepsilon}
\renewcommand{\vec}[1]{\mbox{\boldmath $#1$}}

\renewcommand{\Im}{{\rm Im}\,}
\newcommand{\gMq}{g_{\rm Mqq}}

\newcommand{\ThermalInt}[2]{T\sum_{#1\in\mathds{Z}}\int\frac{\dint^3 #2}{(2\pi)^3}}

\begin{document}
\title{Shear viscosity from Kubo formalism: NJL-model study}

\author{Robert Lang}
  \email{robert.lang@ph.tum.de}
  \affiliation{Physik Department, Technische Universit\"{a}t M\"{u}nchen, D-85747 Garching, Germany}
  \affiliation{Theoretical Research Division, Nishina Center, RIKEN, Wako 351-0198, Japan}
\author{Wolfram Weise}
  \email{weise@ectstar.eu}
  \affiliation{ECT*, Villa Tambosi, I-38123 Villazzano (TN), Italy}
  \affiliation{Physik Department, Technische Universit\"{a}t M\"{u}nchen, D-85747 Garching, Germany}

\date{March 26, 2014}

\begin{abstract}
A large-$N_{\rm c}$ expansion is combined with the Kubo formalism to study the shear viscosity $\eta$ of strongly interacting matter in the two-flavor NJL model. We discuss analytical and numerical approaches to $\eta$ and investigate systematically its strong dependence on the spectral width and the momentum-space cutoff. Thermal effects on the constituent quark mass from spontaneous chiral symmetry breaking are included. The ratio $\eta/s$ and its thermal dependence are derived for different parameterizations of the spectral width and for an explicit one-loop calculation including mesonic modes within the NJL model.
\end{abstract}



\maketitle

\section{Introduction}
Heavy-ion collisions at RHIC \cite{BRAHMSatRHIC05,PHENIXatRHIC05,STARatRHIC05,PHOBOSatRHIC05} and at the LHC \cite{AamodtALICE10,AamodtALICE11,CaffarriALICE12} explore strongly interacting matter under extreme conditions. The quark-gluon matter created in such collisions at temperatures exceeding $T_{\rm c}\approx 0.2$ GeV leaves its indirect signatures in the produced particles at lower temperatures long after thermalization. Transport properties such as the shear viscosity $\eta$ of the highly excited matter are of prime interest in this context. Inertial anisotropies in the collision plane translate into non-trivial particle flow patterns \cite{DerendarzATLAS2013,RoyMohantyChaudhuri2013,DuslingTeaney2008,LuzumGombeaudOllitrault2010}. In particular the elliptic-flow parameter, $v_2$, features a strong dependence on the ratio of shear viscosity to entropy density, $\eta/s$, of the dissipative quark-gluon matter formed in the collision. It is known that the temperature dependence of $\eta/s$ is crucial in 
order to describe the elliptic flow of hadrons in ultrarelativistic heavy-ion collisions at RHIC and LHC \cite{Niemi2011}. At the very high LHC energies, the temperature dependence of $\eta/s$ for $T>T_{\rm c}$ becomes dominant for the elliptic flow.

Two basic approaches are commonly used to deal with non-equilibrium systems and transport properties: the Boltzmann equation \cite{HuiDefuLic2006,SasakiRedlich2010,KhvorostukhinToneevVoskresensky2011} (most frequently applied in relaxation-time approximation) and the Kubo formalism using retarded correlators of the energy-momentum tensor \cite{Alberico2008,HidakaKunihiro2011,HidakaKunihiro2011NJL,NamKao2013}. When examined in comparison, the kinetic approach seems generally to underestimate the shear viscosity \cite{PlumariEtAl2012,PlumariEtAl2013}. The calculation in relaxation-time approximation makes use of the thermal cross sections of the colliding particles, whereas the Kubo formalism asks for the spectral functions of the basic degrees of freedom. These two approaches are connected via the optical theorem. In the present paper we choose the Kubo formalism, realizing at the same time that, in this approach, a perturbative treatment of transport coefficients is insufficient \cite{JeonSkeleton1995,
JeonYaffeSkeleton1996} and requires resummation techniques even in a weak-coupling situation.

We use the Nambu--Jona-Lasinio (NJL) model \cite{Nambu1961tp,Nambu1961,Klevansky1992,KLVW1990,KLVW1990:2,VoglWeise1991,BuballaHabil2005} as a schematic, non-perturbative approach to the thermodynamics of quark matter. Gluonic degrees of freedom are integrated out and hidden in point vertices of the effective interaction between quarks, while all relevant chiral and flavor symmetries and symmetry-breaking patterns of QCD are taken properly into account. The applicability of such a model is supposed to cover a temperature range $T_{\rm c} \lesssim T < \Lambda$, where $T_{\rm c}\approx 0.2$ GeV is the transition temperature from the hadronic to the quark phase and $\Lambda\approx 0.6\;{\rm GeV}$ is the characteristic NJL cutoff scale. We combine the NJL model with a large-$\Nc$ expansion \cite{tHooft1974,QuackKlevansky94,BuballaMuellerWambach2010} to study the shear viscosity of quark matter.

In Section \ref{Sec:EtaLO} the shear viscosity $\eta$ in leading order is deduced. We follow mainly the developments in \cite{Fukutome2006,Fukutome2008Nucl,Fukutome2008Prog} but do not restrict ourselves to the chiral limit and arrive at results under more general assumptions. Taking only the dominant scalar and pseudoscalar channels into account, an infinite number of ring diagrams reduces to just one single generic diagram. Corrections to correlation functions by ladder diagrams are suppressed in a large-$\Nc$ expansion.
However, for the shear viscosity itself a resummation of these subleading diagrams is potentially important, depending on the $\Nc$ scaling of the spectral width \cite{HidakaKunihiro2011,HidakaKunihiro2011NJL}.  This effect has been studied first in \cite{JeonSkeleton1995} for a bosonic field theory.
In the present work resummations in the Kubo sector will not be included. A discussion concerning the 
conditions under which such resummations are necessary will however be given.

The general derivation is followed by a detailed parameter study in Section \ref{Sec:ParStud}: for  $\eta[\Gamma]$ as a functional of the quasiparticle spectral width $\Gamma(p)$ of the quarks. First an analytical result is derived assuming a constant spectral width to start with. Furthermore, the implementation of different parameterizations for $\Gamma(p)$ teaches us about the general dependence of the shear viscosity on the spectral width for a variety of examples. A strong dependence on the pertinent momentum-space cutoff, $\Lambda$, is found, reflecting the sensitivity to physical scales: the characteristic cutoff fixed by the NJL gap equation excludes up to $90\%$ of the mathematically accessible high-momentum contributions to $\eta$, a feature that actually turns out to be a prerequisite for achieving physically meaningful results within this framework. We also investigate to what extent the functional $\eta[\Gamma(p)]$ can be treated perturbatively by expanding in a Laurent series for a ``small'' 
spectral width and comparing with the full result. The impact of thermal constituent quark masses on the shear viscosity is investigated in Section \ref{Sec:ThermoMass}. The thermal quark masses are generated dynamically by the spontaneous chiral symmetry breaking mechanism through the NJL gap equation. Not surprisingly, we find that thermal effects are crucial to obtain physically relevant results for the shear viscosity.

In Section \ref{Sec:GammaNJL} an explicit calculation of the spectral width is performed within the NJL model, using  the one-loop mesonic contributions to the quark self-energy at next-to-leading order in the large-$\Nc$ expansion. Two different physical effects contribute to this width: Landau damping and mesonic recombination. For temperatures well above the critical/crossover temperature the resulting spectral width decreases, implying an increasing shear viscosity $\eta(T)$ in this temperature range.

\section{Shear viscosity at leading order} \label{Sec:EtaLO}
In this work we model quark matter starting from the two-flavor NJL Lagrangian
\begin{equation}
  \label{NJL2}
  \mathcal{L}=\psib \left(\i\dslash-\hat{m}\right)\psi+\frac G2\left[(\psib\psi)^2+(\psib \i\gamma_5\vec{\tau}\psi)^2\right],
\end{equation}
where $\psi = (u,d)^{\rm T}$ is the isospin doublet quark field, $\hat{m} = {\rm diag}(m_{\rm u},m_{\rm d})$ is the current quark mass matrix (we work in the isospin limit, $m_{\rm u}= m_{\rm d} \equiv m$), $G$ denotes the scalar/pseudoscalar coupling, and $\vec{\tau}$ collects the three Pauli isospin matrices. Vector or axialvector terms are not considered in this work. The large masses of the corresponding quark-antiquark modes make their contributions to the relevant correlation functions far less important than those of pseudoscalar and scalar modes. 

In the Kubo formalism \cite{Kubo57} transport coefficients are related to retarded correlators of energy-momentum tensors, i.e. to four-point functions in Matsubara space. The energy momentum tensor of the NJL model is simply
\begin{equation}
  \label{Tmunu}
T_{\mu\nu}=\i\psib\gamma_\mu\partial_\nu\psi-g_{\mu\nu}\mathcal{L}
\end{equation}
in terms of the quark fields $\psi$. The Kubo formula for the shear viscosity reads
\begin{equation}
\begin{aligned}
  \label{KuboEta}
\eta(\omega) =
  \frac{\beta}{15}\int_0^\infty\dint t\;\e^{\i\omega t}\int\dint^3r\; (T_{\mu\nu}(\vec{r},t),T^{\mu\nu}(0))\;,
\end{aligned}
\end{equation}
with $\beta = 1/T$ the inverse temperature.
The correlator $(X,Y)$ is defined by\footnote{Due to $\e^{-\beta H}X(t)\e^{\beta H}=X(t+\i\beta)$, which implies $\langle X(t)Y(t'+\i\beta)\rangle=\langle Y(t')X(t)\rangle$, it follows that this correlator is symmetric in its arguments: $(X,Y)=(Y,X)$.}
\begin{equation}
  (X,Y)=T\int_0^\beta\dint\xi\;\langle\e^{\xi H}X\e^{-\xi H}Y\rangle\;,
\end{equation}
where $H$ is the NJL Hamiltonian and $\langle\mathcal{\cdot}\rangle=\Tr\left(\cdot\,\e^{-\beta H}\right)$ denotes the thermal expectation value. An equivalent reduced expression for the shear viscosity is also frequently used in the literature \cite{Alberico2008}:
\begin{equation}
\label{KuboEtaAlternative}
  \eta(\omega)=\beta\int_0^\infty\dint t\;\e^{\i\omega t}\int\dint^3r\; (T_{21}(\vec{r},t),T_{21}(0))\;,
\end{equation}
written in terms of only one component of the energy-momentum tensor. The relative factor $15$ in comparison with Eq.\,\eqref{KuboEta} results from the following identity ($i,j\in\{1,2,3\}$ with $i\neq j$):
\begin{equation}
  \int\dint^3x\; x_i^2\,x_j^2\,f(x^2)=\frac{1}{15}\int\dint^3x\;x^4f(x^2)\;.
\end{equation}
The (classical) components of the energy-momentum tensor are real quantities, $T_{\mu\nu}\in\mathds{R}$. It follows that the \textit{static} shear viscosity $\eta(\omega = 0)$ is also real:
\begin{equation}
  \eta(\omega)^*=\eta(-\omega)\;\;\;\Rightarrow\;\;\;\eta:=\eta(0)\in\mathds{R}\;.
\end{equation}
Neglecting surface terms at infinite time, one derives
\begin{equation}
\label{DefRetGreenFctFromDifference}
  \eta(\omega)=\frac{\i}{\omega}\left[\Pi^{\rm R}(\omega)-\Pi^{\rm R}(0)\right],
\end{equation}
with the retarded correlation function
\begin{equation}
  \label{DefPiROmega}
\Pi^{\rm R}(\omega)=-\i\int_0^\infty\dint t\;\e^{\i\omega t}\int\dint^3r\,\langle[T_{21}(\vec{r},t),T_{21}(0)]\rangle\;.
\end{equation}
The static shear viscosity ($\omega\to 0$) follows as
\begin{equation}
  \eta=-\left.\frac{\dint}{\dint\omega}\,\Im\Pi^{\rm R}(\omega)\right|_{\omega=0}\;.
\end{equation}
The calculation of the retarded correlator can be performed switching to the Matsubara formalism and calculating 
\begin{equation}
  \Pi(\omega_n)=\int_0^\beta\dint\tau\;\e^{\i\omega_n\tau}\int\dint^3 r\; \langle \mathcal{T}_\tau
\big(T_{21}(\vec{r},\tau)T_{21}(0)\big)\rangle\;,
\end{equation}
where have applied a Wick rotation $\tau=\i t$ and introduced the time-ordering symbol in imaginary time, $\mathcal{T}_\tau$. Note that whereas the underlying (quark) degrees of freedom are fermionic, the Matsubara frequencies relevant for the correlator $\Pi$ are bosonic, $\omega_n=2\pi n T$, since the fermion fields under the integral group together to form quantities of bosonic character: $\psib(\cdot)\psi$. The global sign of $\Pi(\omega_n)$ is fixed by the sign convention for analytical continuations:
\begin{equation}
  \left.\Pi(\omega_n)\right|_{i\omega_n=\omega\pm\i\eps}= -\,\Pi^{\rm R/A}(\omega)\;,
\end{equation}
where the upper and lower sign in $\pm\i\eps$ corresponds to the retarded and advanced correlation function, respectively.
The correlator $\Pi(\omega_n)$ is governed by non-perturbative physics resulting from the underlying interactions of the NJL model. We now apply a large-$\Nc$ expansion and organize this correlator in ring diagrams, ladder diagrams and higher-order terms:
\begin{equation}
  \label{KuboNcExpansionFourPointFunction}
  \begin{minipage}{0.08\textwidth} \mbox{$\Pi(\omega_n)=$} \end{minipage}  
  \begin{minipage}{0.15\textwidth}
    \includegraphics[width=\textwidth]{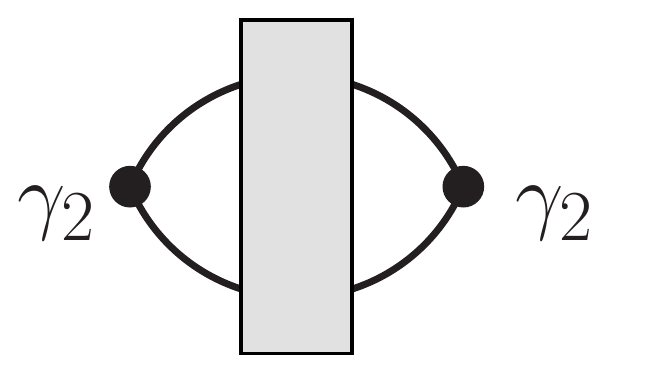}
  \end{minipage}
  \hspace{-0.2cm} \begin{minipage}{0.2\textwidth}
  \mbox{$= \mathcal{O}(N_{\rm c}^1)+\mathcal{O}(N_{\rm c}^0)+ \ldots $}
\end{minipage}
\end{equation}
The four-point coupling of the NJL Lagrangian \eqref{NJL2} effectively incorporates gluonic degrees of freedom resulting in the scaling $G\sim 1/\Nc$. At leading order $\mathcal{O}(N_{\rm c}^1)$ there is just a one-loop diagram contributing to the four-point correlator, given that the NJL Lagrangian in its simplest form \eqref{NJL2} takes into account only scalar and pseudoscalar interactions: $\Gamma\in\{\Unit,\i\gamma_5\}$. Iterating these interaction kernels in ring diagrams at leading order in $1/\Nc$ to $\Pi(\omega_n)$ does not affect the correlator:
\begin{equation}
\label{Chain}
\hspace{0.5cm} \begin{minipage}{0.04\textwidth}
\end{minipage}
  \begin{minipage}{0.3\textwidth}
    \includegraphics[width=\textwidth]{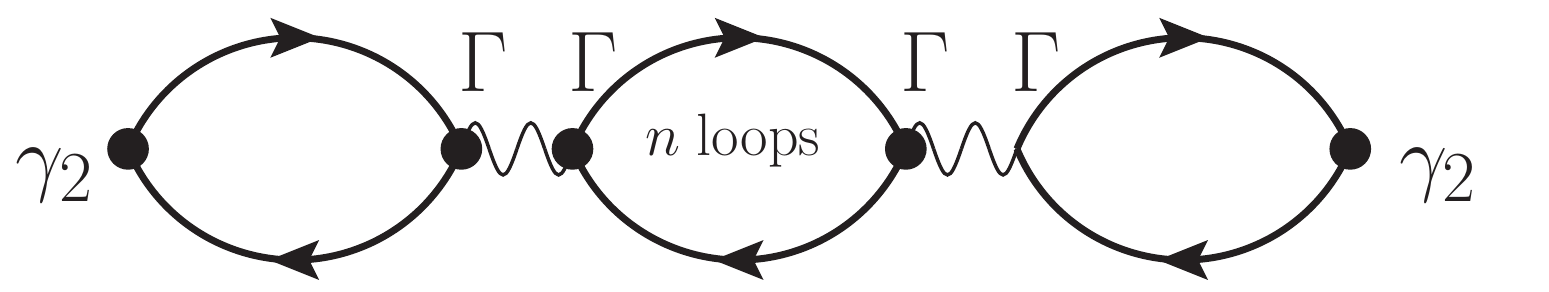}
  \end{minipage}
\begin{minipage}{0.05\textwidth} \!=\;0\;, \end{minipage}
\end{equation}
because the trace (in momentum and Dirac space) in the first ring vanishes due to the orthogonal operator structure involving the combination of $\gamma_2$ and $\Gamma$:
\begin{equation}
\begin{aligned}
  \label{KuboRingAZero1}
  &\ThermalInt{n}{p}\; \Tr\left[\gamma_2\, G_\beta(\vec{p},\nu_n)\, \Gamma\, G_\beta(\vec{p},\nu_n)\right] =\\
  &  =\ThermalInt{n}{p}\; \frac{1}{(\nu_n^2+\vec{p}^2+M^2)^2}\; \\
  &\hspace{0.5cm} \times\, \Tr\left[\gamma_2\Gamma M^2 +\gamma_2\slashed{p}\Gamma\slashed{p}+\gamma_2\slashed{p}\Gamma M+\gamma_2\Gamma\slashed{p} M\right]=0\;,
\end{aligned}
\end{equation}
where we have used the notation $\slashed{p}=\nu_n\gamma_4-\vec{p}\cdot\vec{\gamma}$ and the full Matsubara propagator
\begin{equation}
  G_\beta(\vec{p},\nu_n)=\frac{\slashed{p}+M}{\nu_n^2+\vec{p}^2+M^2}\;,
\end{equation}
with frequencies $\nu_n=(2n+1)\pi T-\i\mu$. Exchange (ladder diagram) corrections to the chain in Eq.~\eqref{Chain} are non-vanishing but of subleading order in $1/\Nc$, because each rank in the ladder gives rise to a suppression factor $G^2\Nc\sim 1/\Nc$. Note that adding one rank introduces two additional momentum integrations but only one additional color trace.

The shear viscosity in the NJL model has been deduced previously in Refs. \cite{Fukutome2006,Fukutome2008Nucl,Fukutome2008Prog} using the Kubo formula, but assuming the quarks to be in the chiral limit, $m=0$. We point out that this result can in fact be derived without assuming to work in the chiral limit. Setting the current quark masses to zero is \textit{not necessary} to ensure the absence of iterated ring-diagram contributions when taking only scalar and pseudoscalar interactions of the NJL model into account. Iterated ring diagrams involving these interactions vanish naturally. (Note that even in the chiral limit and the Nambu-Goldstone phase, the second term of the trace in Eq.~\eqref{KuboRingAZero1}, $\Tr\left[\gamma_2\slashed{p}\Gamma\slashed{p}\right]$ would survive in the presence of vector interactions, but their contribution to the correlators would be small as mentioned before).

Collecting all arguments, we can summarize in general: for purely fermionic theories $\mathcal{L}=\mathcal{L}_{\rm kin}+\mathcal{L}_{\rm int}$ with momentum-independent pseudoscalar/scalar interactions and $2n$-vertices that scale as $G_{2n}\sim 1/\Nc^{n-1}$, the dominant contribution to the correlation function $\Pi^{\rm R}(\omega)$ in Matsubara space is:
\begin{equation}
\label{PiRingLO}
  \begin{minipage}{0.08\textwidth} \mbox{$\Pi(\omega_n)=$} \end{minipage}  
  \begin{minipage}{0.125\textwidth}
    \includegraphics[width=\textwidth]{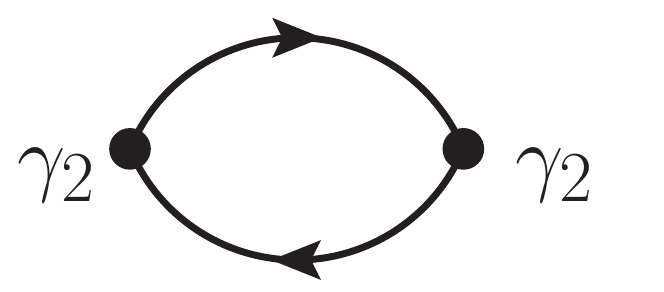}
  \end{minipage}
  \begin{minipage}{0.1\textwidth} \mbox{$\;+\;\mathcal{O}(\Nc^0)\;.$} \end{minipage}
\end{equation}
With the definition of the spectral function,
\begin{equation}
  \rho(\omega,\vec{p})=-\frac{1}{\pi}\Im G^{\rm R}(\omega,\vec{p})=\frac{1}{2\pi\i}\left(G^{\rm A}(\omega,\vec{p})-G^{\rm R}(\omega,\vec{p})\right),
\end{equation}
and using residue calculus one derives:
\begin{equation}
\begin{aligned}
\label{EtaLeadingNcAsTwoSpectralDensityPap}
  \eta &={\pi\over T} \int_{-\infty}^\infty \dint\eps\int\frac{\dint^3 p}{(2\pi)^3}\;p_x^2\, n_{\rm F}^+(\eps)\big(1-n_{\rm F}^+(\eps)\big) \\
  &\hspace{2.5cm} \times\,\Tr\left[\gamma_2\,\rho(\eps,\vec{p})\,\gamma_2\,\rho(\eps,\vec{p})\right],
\end{aligned}
\end{equation}
with the Fermi-Dirac distribution
\begin{equation}
n_{\rm F}^+(E)=\frac{1}{1+\e^{\beta (E-\mu)}}\;.
\end{equation}
As in \cite{Fukutome2006} the dressed quark propagator is written as
\begin{equation}
  \label{KuboQuasiPartApprQuarkProp}
G^{\rm R/A}(p_0,\vec{p})=\frac{1}{\slashed{p}-M\pm\i\,{\rm sgn}(p_0)\Gamma(p)}\;,
\end{equation}
with the quasiparticle mass $M$ and width $\Gamma(p)$. The next step is to relate this spectral width to the shear viscosity $\eta$. Even in the chiral limit the dynamical NJL mechanism of spontaneous chiral symmetry breaking generates a large constituent quark mass in the vacuum: $M\approx 0.3\;{\rm GeV}$, see the brief discussion in Section \ref{Sec:ThermoMass}. Apart from this mechanism, the thermal environment at temperature $T$ and baryo-chemical potential $\mu$ of the quarks affects parameterically both the dynamical quark mass $M(T,\mu)$ and the spectral width $\Gamma(p;T,\mu)$.

\begin{figure}[t!]
\begin{center}
  \subfigure[Temperature dependence of the shear viscosity at vanishing chemical potential]{\includegraphics[width=0.48\textwidth]{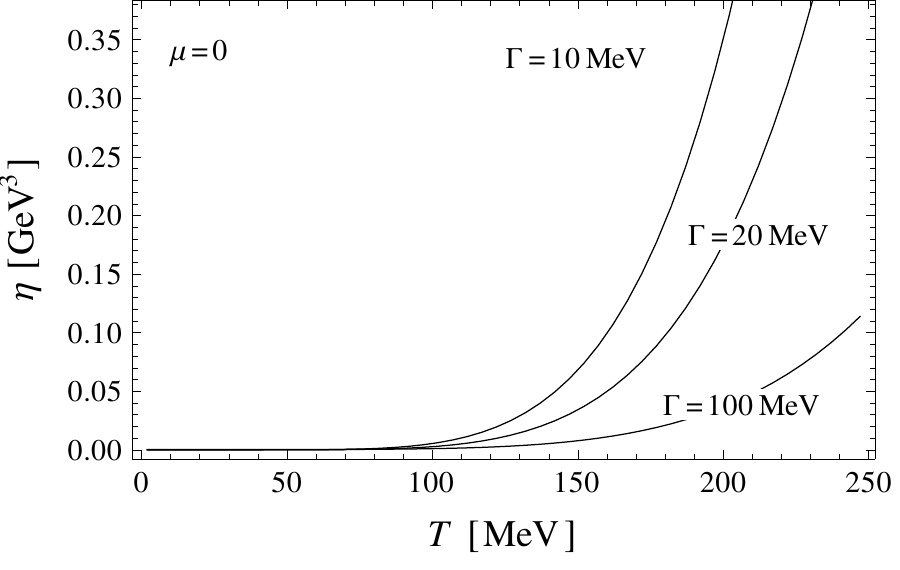}}
  \subfigure[Dependence of the shear viscosity on the chemical potential]{\includegraphics[width=0.48\textwidth]{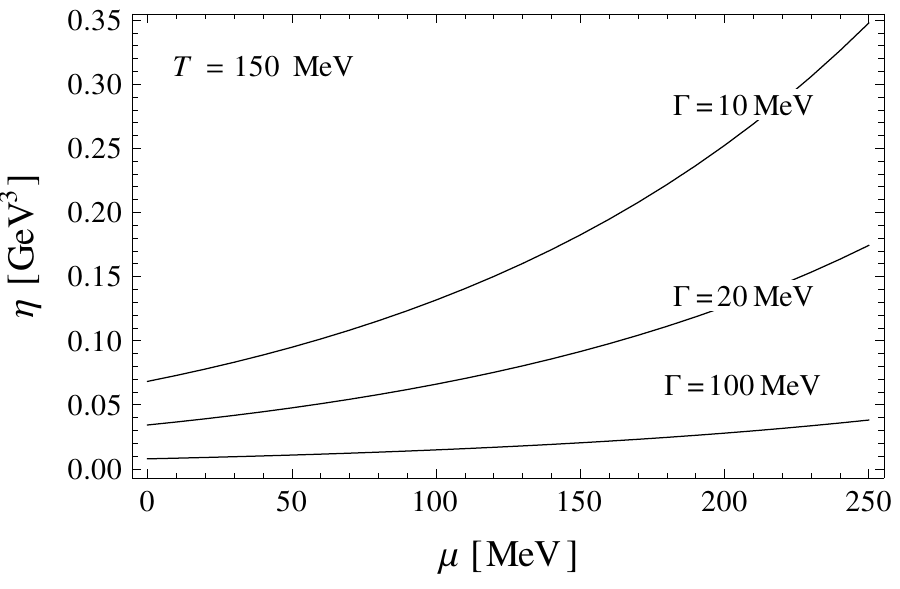}}
\caption{Shear viscosity $\eta$ for constant spectral width $\Gamma$ as function of temperature $T$ and quark chemical potential $\mu$. In this analytical calculation based on Eq.~\eqref{KuboAnalyitcalPIntegral} no momentum cutoff has been used. We have chosen $M=100\;{\rm MeV}$ for in order to compare our results to those in \cite{Fukutome2006}.}
\label{Fig1}
\end{center}  
\end{figure}

The spectral function $\rho$ is represented in the standard form of a generalized Breit-Wigner shape, to be inserted in Eq.\,\eqref{EtaLeadingNcAsTwoSpectralDensityPap}:
\begin{equation}
\begin{aligned}
\label{EtaLeadingNcAsSpectralWidth}
  \eta[\Gamma(p)] &=\frac{16\Nc\Nf}{15\pi^3 T}\int_{-\infty}^\infty\dint\eps\int_0^\infty\dint p\,p^6 \\
  &\times\,\frac{M^2\,\Gamma^2(p)\,n_{\rm F}^+(\eps)(1-n_{\rm F}^+(\eps))}{\left[(\eps^2-p^2-M^2+\Gamma^2(p))^2+4M^2\Gamma^2(p)\right]^2}\;.
\end{aligned}
\end{equation}
Even though we started from the NJL model in our derivation, the expression \eqref{EtaLeadingNcAsSpectralWidth} is generic for a system of strongly interacting Fermions, with real and imaginary parts of their self-energies encoded in $M$ and $\Gamma$, respectively.
In general, the non-perturbative origin of $\Gamma(p)$ does not permit expanding the functional $\eta[\Gamma]$ in a Laurent series. In perturbative approaches (e.g. in chiral perturbation theory at low temperatures) such a treatment is possible: $\eta[\Gamma]\sim 1/\Gamma$ \cite{LangKaiserWeise2012}.

In this context we comment briefly on the issue of ladder resummation and its correction to Eq.~\eqref{PiRingLO}. If one assumes the spectral width $\Gamma\sim 1/\Nc$ to be suppressed for large $\Nc$ as suggested by hot-QCD calculations, then the superficial $\Nc$ counting of Eqs.~\eqref{EtaLeadingNcAsTwoSpectralDensityPap} and \eqref{EtaLeadingNcAsSpectralWidth}, $\eta\sim\Nc$, is spoiled: in this case, the integrand becomes highly singular in the large-$\Nc$ limit (``pinch poles'' as described in \cite{JeonSkeleton1995,HidakaKunihiro2011}), resulting in an additional factor $\Nc$, therefore $\eta\sim\Nc^2$. We find in the limit $\Nc\to\infty$, i.e. $\Gamma\to 0$:
\begin{equation}
\label{EtaSmallGamma}
  \eta[\Gamma]\to\frac{2\Nc\Nf}{15\pi^2 T}\int_{|\eps|>M}\dint\eps\,\frac{(\eps^2-M^2)^{5/2}\,n_{\rm F}^+(\eps)(1-n_{\rm F}^+(\eps))}{M\,\Gamma(\sqrt{\eps^2-M^2})}\;.
\end{equation}
In contrast to the non-perturbative result in Eq.~\eqref{EtaLeadingNcAsSpectralWidth} the $\eps$-integration excludes the region $|\eps|<M$. This is due to the delta functions appearing in the limit of small $\Gamma$. The momentum integration of the integrand involving $\delta(\eps^2-p^2-M^2)$ is readily carried out.

\section{Exploratory studies of the shear viscosity} \label{Sec:ParStud}
\subsection{Analytical results} 
Consider now first the case of a constant spectral width, $\Gamma={\rm const}$. This rough schematic approximation allows for an analytical treatment of the momentum integral in Eq.~\eqref{EtaLeadingNcAsSpectralWidth} and one is left with the numerical $\eps$-integration only. For the momentum integral we use the following identity:
\begin{equation}
\begin{aligned}
\label{KuboAnalyitcalPIntegral}
  \int_0^\infty &\dint p\;\frac{p^6}{[(A-p^2)^2+B^2]^2}= \\
  &\hspace{-0.5cm}=\frac{\pi}{8\sqrt{2}}\frac{\sqrt{\sqrt{A^2+B^2}-A}}{B^4} \left[(2A^2+3B^2)\sqrt{A^2+B^2} \right.\\
  &\hspace{3.7cm} \left. +2A(A^2+2B^2)\right],
\end{aligned}
\end{equation}
where we have introduced $A=\eps^2-M^2+\Gamma^2$ and $B=2M\Gamma$. This result is found by extending the integration region to negative $p$ (the integrand is an even function of $p$) and using residue calculus. Having performed the $p$-integration analytically reduces the computation time by roughly one order of magnitude. Furthermore, it helps finding an appropriate approximation scheme for the whole $(\eps,p)$-integration when the spectral width is momentum dependent.

Fig.~\ref{Fig1} shows the results for $\eta$ assuming $\Gamma={\rm const.}$ For $\Gamma\to 0$ the shear viscosity diverges, as it follows from Eq.~\eqref{EtaSmallGamma}. This limit describes a system of free quarks for which the mean free path is infinite. With increasing temperature and chemical potential, the shear viscosity increases, but the dependence on temperature is more pronounced. Compare these figures to those in Ref.~\cite{Fukutome2006}, where $\eta(\Gamma)$ has been evaluated numerically without a momentum-space cutoff, equivalent to our analytical approach based on Eq.~\eqref{KuboAnalyitcalPIntegral}.

Inspecting the detailed behavior of the integrand in Eq.~\eqref{EtaLeadingNcAsSpectralWidth}, a convergence criterion for the shear viscosity in the absence of a momentum-space cutoff can be derived: 
\begin{quote}
In order for the shear viscosity $\eta[\Gamma]$ as functional of $\Gamma(p)$ to be convergent, the asymptotic $\Gamma(p)$ should not converge too rapidly to zero:
\begin{equation}
\label{CriterionEtaGamma}
  \eta[\Gamma(p)]<\infty \;\;\; \Leftrightarrow \;\;\; p^3\e^{-\beta p/2}\in {\rm o}(\Gamma(p))\;,
\end{equation}
\end{quote}
where ${\rm o}(\cdot)$ denotes the little Landau symbol\footnote{The notation $f\in{\rm o}(g)$ is used to express accurately that ``$f$ is growing less fast than $g$'', meaning that $f(x)/g(x)\to 0$ for $x\to\infty$. More intuitively, this also means that ``$g$ grows much faster than $f$''.}. Possible parameterizations of $\Gamma(p)$ satisfying this constraint are:
\begin{equation}
\begin{aligned}
  \label{ModelsMomentumDepGamma}
  {\rm constant:}~~~~~\Gamma_{\rm const} &= 100\;{\rm MeV}\;,\\
  {\rm exponential:}~~~\Gamma_{\rm exp}(p) &= \Gamma_{\rm const}\,\e^{-\beta p/8}\;,\\
  {\rm Lorentzian:}~~~\Gamma_{\rm Lor}(p) &= \Gamma_{\rm const}\,\frac{\beta p}{1+(\beta p)^2}\;,\\
  {\rm divergent:}~~~\Gamma_{\rm div}(p) &= \Gamma_{\rm const}\,\sqrt{\beta p}\;.
\end{aligned}
\end{equation}
Note that all these parameterizations lead to a finite shear viscosity and no mathematical regularization must be applied, compare the cutoff discussion in Section \ref{SectionCutOff}. The particular shapes of these prototype widths have been chosen because of their different behavior at small and large momenta: vanishing or non-vanishing $\Gamma(p=0)$, convergent or divergent $\Gamma(p)$ for $p\to\infty$. These prototypes represent physical spectral widths in several theories \cite{LangKaiserWeise2012}: $\Gamma(p)$ in $\phi^4$ theory, for instance, is a monotonous function and converges  to zero for large momenta. This can be described by the Lorentz parameterization for large momenta: $\lim_{p\to\infty}\Gamma_{\rm Lor}(p)\sim T/p$. In contrast, the spectral width of an interacting pion gas diverges for $p\to\infty$.

\subsection{Numerical approximation scheme} Our numerical approximation of 
\begin{figure}[t!]
\begin{center}
  \includegraphics[width=0.45\textwidth]{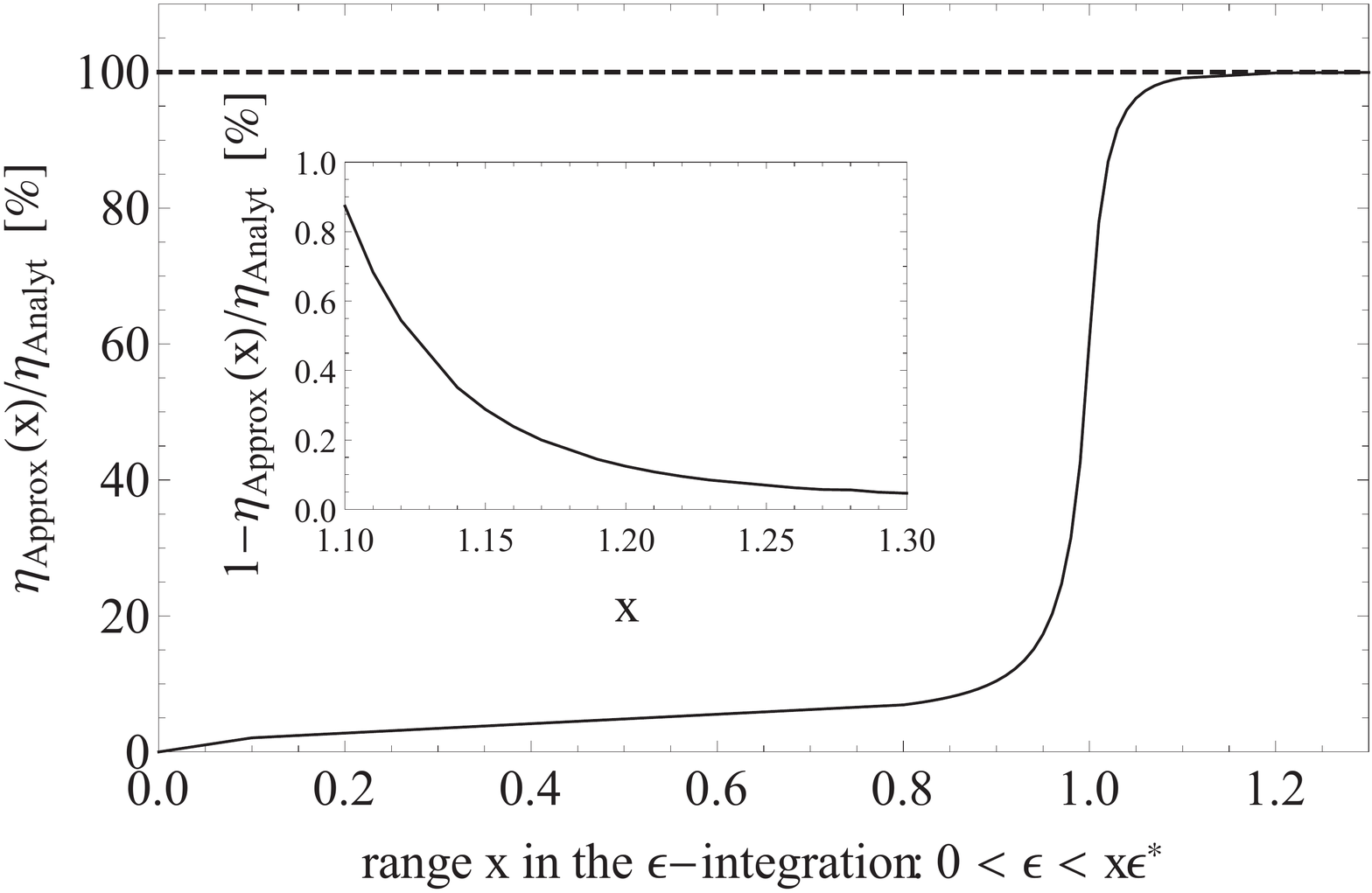}
\caption{Accuracy of the numerical approximation scheme for the $\eps$-integral, Eq.~\eqref{EtaLeadingNcAsSpectralWidth}: to reach an accuracy of $10^{-4}$ it is sufficient to restrict its range to $|\eps(p)|<1.3\,\eps^*(p)$, see Eq.~\eqref{EpsStarMaximizer}.}
\label{Fig2}
\end{center}
\end{figure}
$\eta[\Gamma(p)]$ is based on the observation that its integrand typically ranges over many orders of magnitude. For every momentum $p$ there is a maximum of the integrand in Eq.~\eqref{EtaLeadingNcAsSpectralWidth}, located at the denominator's minimizer
\begin{equation}
\label{EpsStarMaximizer}
  \eps^*(p)=\sqrt{p^2+M^2(T,\mu)-\Gamma^2(p;T,\mu)}\;.
\end{equation}
Adaptive methods do not work when facing a sharp peak structure: either the step size becomes too small for fast convergence (or convergence at all), or the most important contribution in the vicinity of the peak is not sampled by a step size that is too coarse. We overcome this numerical issue by cutting the $\eps$-integration and allowing only $|\eps(p)|<x\eps^*(p)$ for some $x\gtrsim 1$. In comparison with the analytical result for $\Gamma(p)={\rm const.}$ we find that $x=1.3$ is sufficient to produce accurate results within a relative error of $10^{-4}$, see Fig.~\ref{Fig2}.

For the momentum-dependent parameterizations \eqref{ModelsMomentumDepGamma} of $\Gamma(p)$ the integrands for $\eta$ in Eq.~\eqref{EtaLeadingNcAsSpectralWidth} look qualitatively the same as for a constant spectral width.  We therefore expect the described numerical scheme to work well also in these and more physical cases, where full momentum dependence and effects from the thermal environment are (parameterically) taken into account.

\subsection{Cutoff dependence} \label{SectionCutOff}
\begin{figure}[t!]
\begin{center}
  \subfigure[No momentum cutoff applied]{\includegraphics[width=0.46\textwidth]{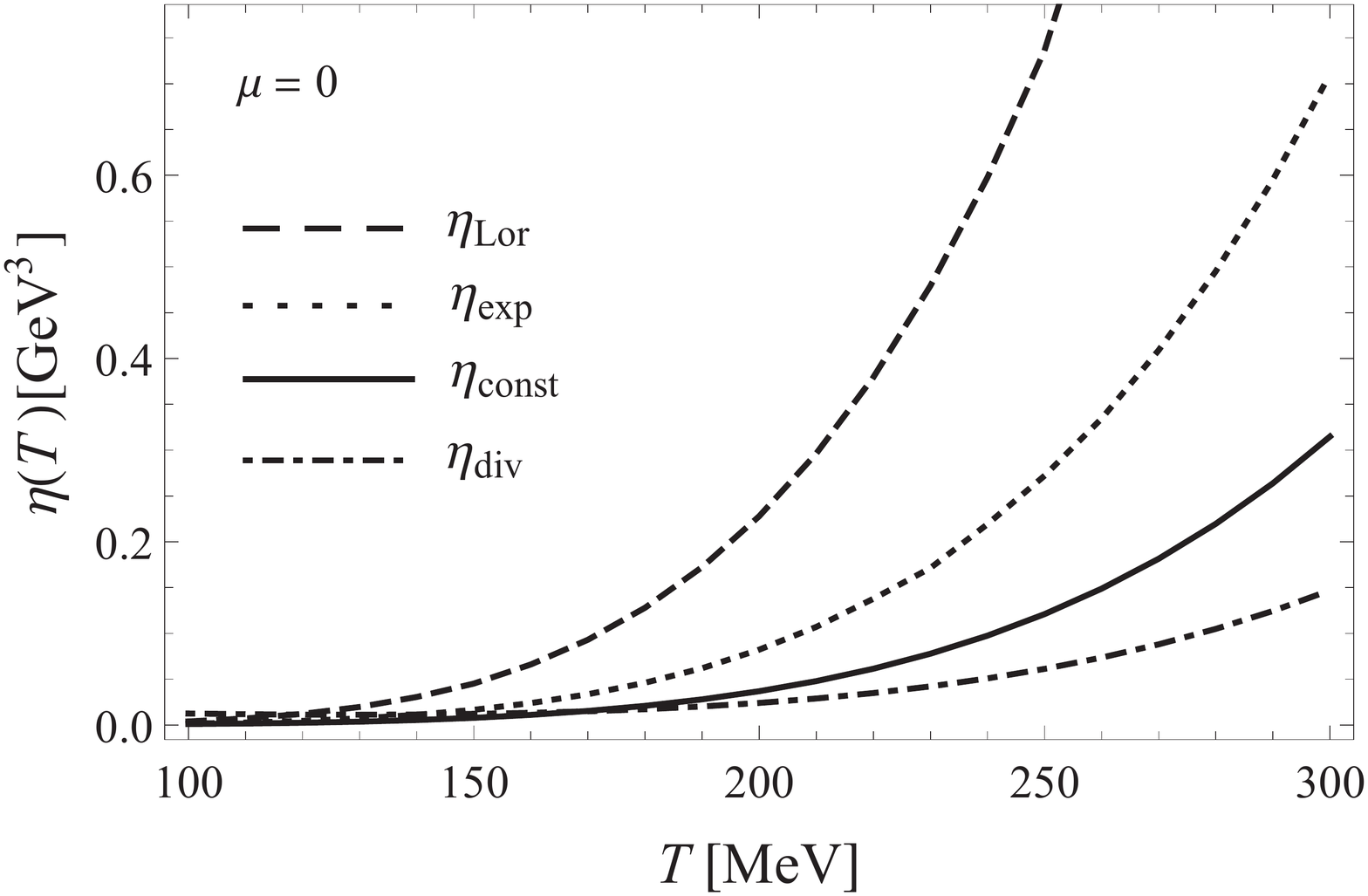}} \hspace{1cm}
  \subfigure[Momenta restricted to $p\leq\Lambda$]{\includegraphics[width=0.46\textwidth]{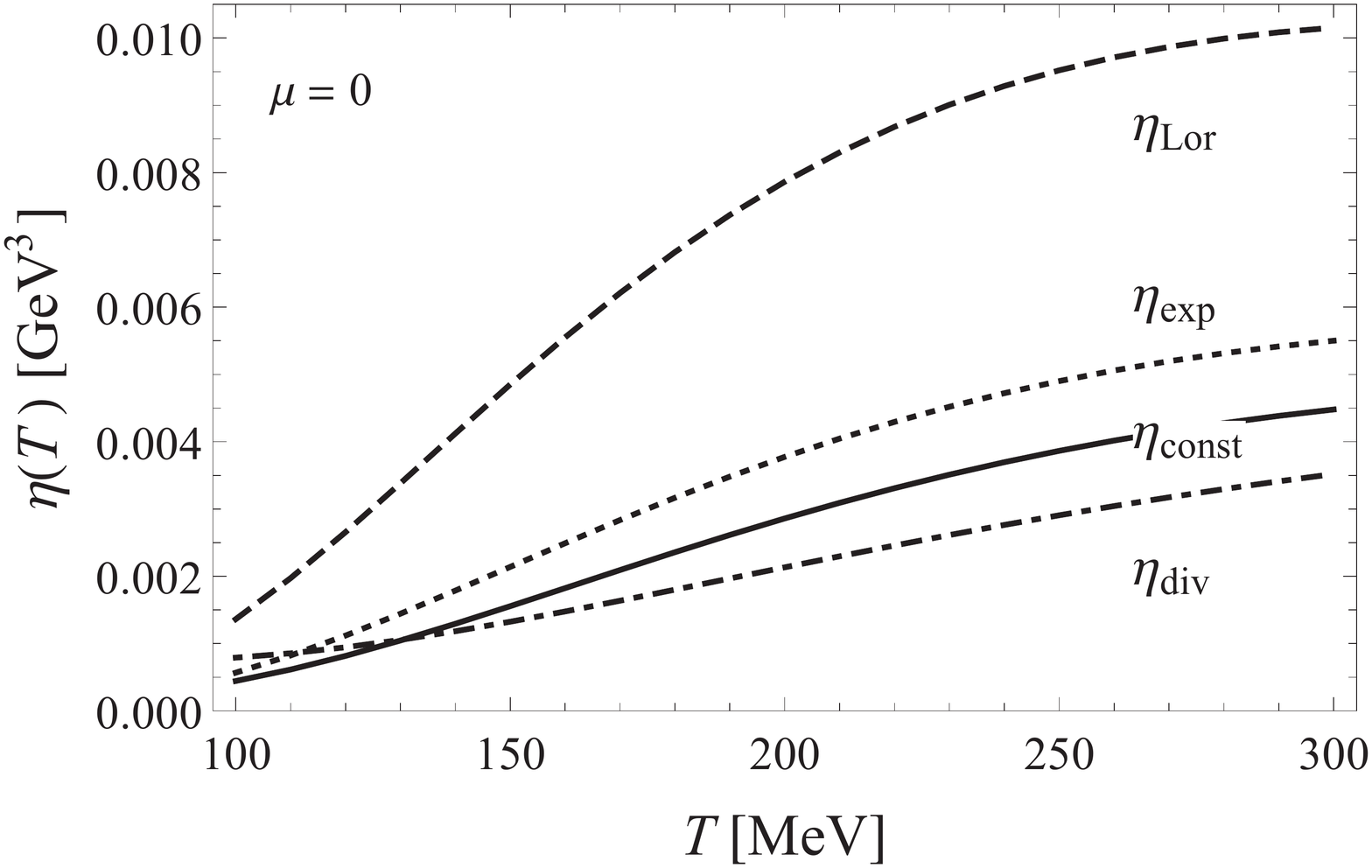}}
\caption{Shear viscosity $\eta$ as function of temperature at vanishing chemical potential, for different schematic parameterizations \eqref{ModelsMomentumDepGamma} of the spectral width $\Gamma(p)$. Sequence of curves and qualitative change from scenarios without (a) and with (b) momentum cutoff $\Lambda=650\,{\rm MeV}$ are discussed in the text.}
\label{Fig3}
\end{center}
\end{figure}

Generally, the shear viscosity increases when the spectral width decreases, compare Eq.~\eqref{EtaSmallGamma}. This behavior is also visible in Fig.~\ref{Fig3}(a) when comparing our different parameterizations of $\Gamma(p)$: the ``more divergent'' the spectral width as $p\rightarrow \infty$, the smaller the corresponding shear viscosity:
\begin{equation}
  \eta_{\rm Lor}>\eta_{\rm exp}>\eta_{\rm const}>\eta_{\rm div}\,,
\end{equation}
using notations as in Eq.~\eqref{ModelsMomentumDepGamma}. This sequence is implied by the corresponding (inverse) order for the spectral widths. These arguments hold also for non-vanishing chemical potentials. Assuming the spectral width to be independent of the chemical potential as in our parameterizations of $\Gamma(p)$ in Eq.~\eqref{ModelsMomentumDepGamma}, the shear viscosity increases for increasing $\mu$, but the qualitative shape of $\eta(T)$ does not change. We note that the results in Fig.~\ref{Fig3} have been derived using a constant constituent quark mass $M=325\;{\rm MeV}$, see the brief discussion in Section \ref{Sec:ThermoMass}.

The integrand of $\eta[\Gamma(p)]$, Eq.~\eqref{EtaLeadingNcAsSpectralWidth}, is sizable for unphysically large momenta, so we expect a strong cutoff dependence. In the NJL model the quasiparticle interactions are restricted to quark momenta $p\le\Lambda=650\;{\rm MeV}$. Quarks with momenta $p>\Lambda$ do not interact and have infinite mean free paths. Retricting the momentum integration to the interval $p \le \Lambda$, we find a shear viscosity as shown in Fig.~\ref{Fig3}(b). Excluding $p>\Lambda$ reduces the shear viscosity by one order of magnitude at low temperatures and even by two orders of magnitude at high  $T$. As expected, this expresses a very strong cutoff dependence. In addition to these quantitative differences, the qualitative behavior of the shear viscosity also changes strongly and flattens for high temperatures.

\begin{figure}[t!]
\begin{center}
  \subfigure[Absolute cutoff dependence; the vertical line represents the position of the physical NJL cutoff, $\Lambda=650\,{\rm MeV}$]{\includegraphics[width=0.46\textwidth]{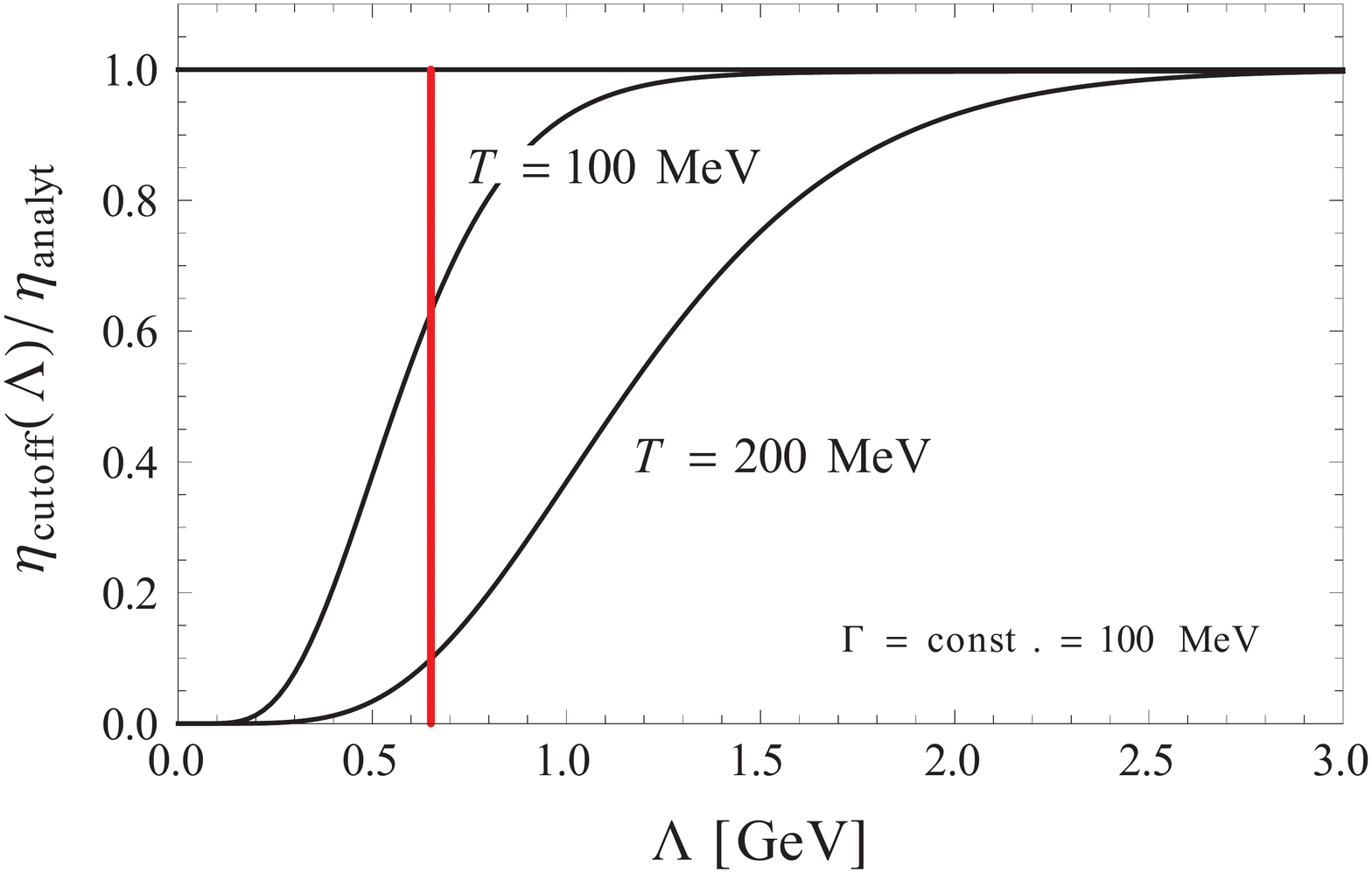}} \hspace{1cm}
  \subfigure[Relative cutoff dependence around $\Lambda=650\,{\rm MeV}$]{\includegraphics[width=0.46\textwidth]{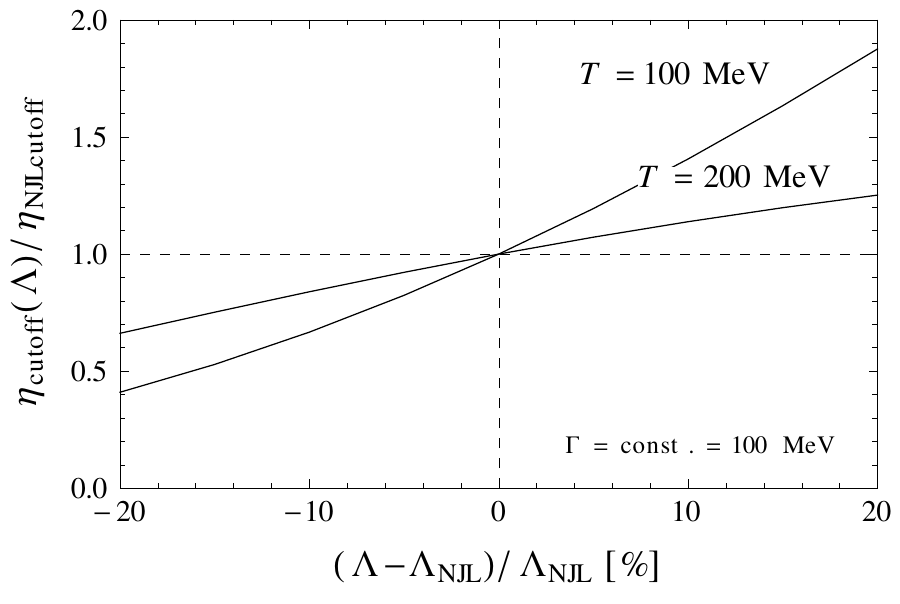}}
\caption{Absolute (a) and relative (b) cutoff dependence of the shear viscosity, demonstrating the suppression of high-momentum contributions when the standard (physical) NJL cutoff is used. The plots are drawn at $\mu=0$ and for constant spectral width $\Gamma=100\;{\rm MeV}$.}
\label{Fig4}
\end{center}
\end{figure}

This strong cutoff dependence is investigated in more detail in Fig.~\ref{Fig4}: the contributions taken into account (compared to the analytical result for $\eta$) depend strongly on temperature and just weakly on the chemical potential. At $T=200\,{\rm MeV}$ the momentum cutoff excludes about $90\%$ of the full integral extended to infinity, see Fig.~\ref{Fig4}(a). As shown in Fig.~\ref{Fig4}(b),  varying the cutoff by up to $\pm 20\%$ implies for $\eta$ a change of up to $100\%$.

To assess the order of magnitude of the NJL shear viscosity, a comparison with $\eta(T)$ for other systems is instructive.  For example, an interacting pion gas treated within the framework of chiral perturbation theory \cite{LangKaiserWeise2012} has a typical shear viscosity of order
$\eta(T)\approx 40\,{\rm MeV}/{\rm fm}^2 \approx 1.6\cdot 10^{-3}\,{\rm GeV}^3$
at $T\approx 100\;{\rm MeV}$. This is a similar order of magnitude as the results shown in Fig.~\ref{Fig3}(b) when applying the NJL cutoff $\Lambda=650\,{\rm MeV}$. We recall that this cutoff is fixed by reproducing physical observables such as the pion decay constant in vacuum. A physically meaningful order of magnitude for $\eta$ then follows naturally.

\subsection{Perturbative aspects of {\boldmath $\eta[\Gamma]$}} \label{Sec:PertEta}
\begin{figure}[t!]
\begin{center}
  \includegraphics[width=0.46\textwidth]{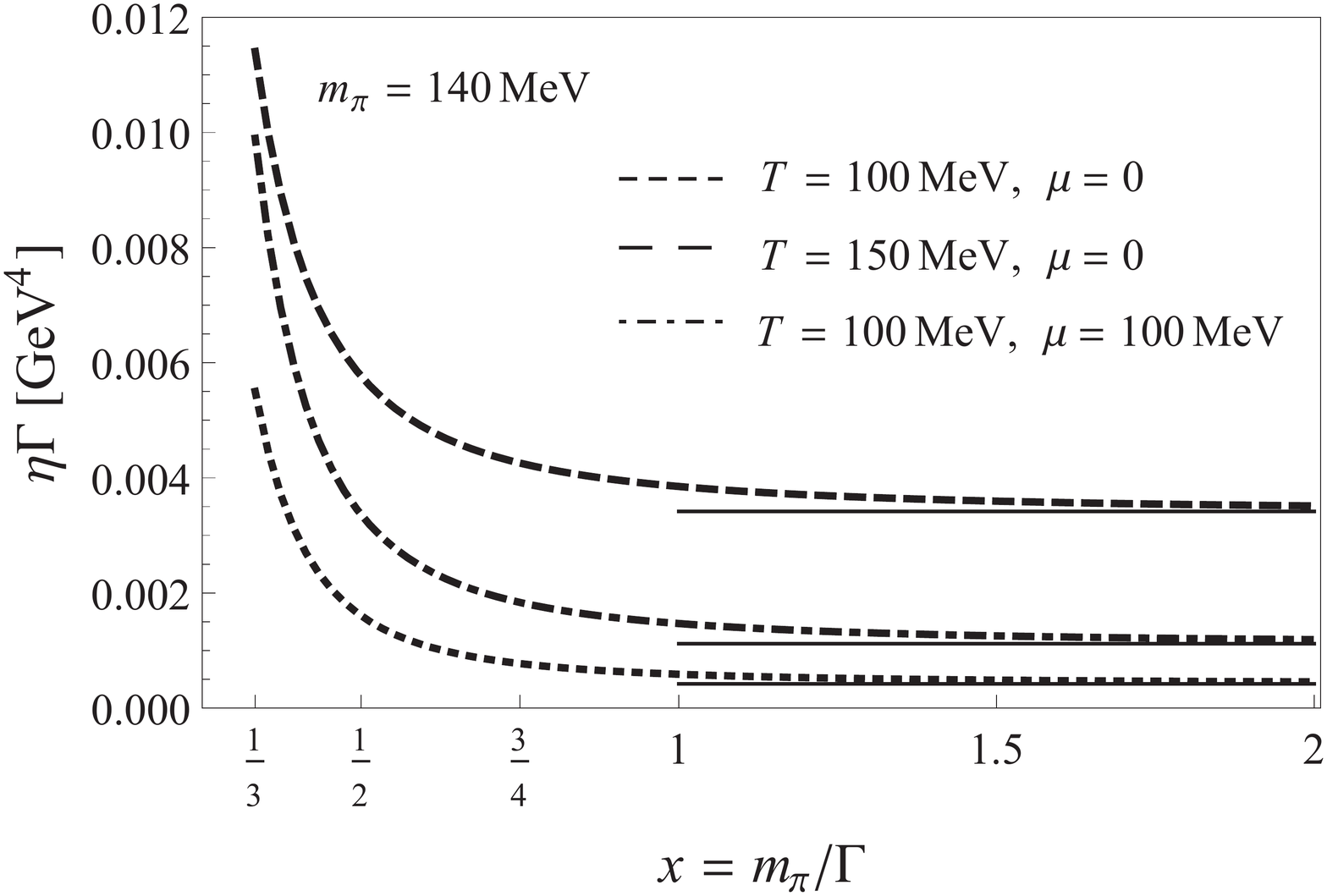}
  \caption{Scaling of $\eta\cdot\Gamma$ for different $T$ and $\mu$ as function of the inverse width expressed in units of the pion mass $m_\pi$. Solid horizontal lines correspond to the residues $A_{-1}$ of $\eta[\Gamma]$ in Eq.~\eqref{LaurentExpEtaGamma}. A constituent quark mass $M=100\;{\rm MeV}$ has been used for convenience.}
  \label{Fig5}
\end{center}
\end{figure}
We have already mentioned that the shear viscosity $\eta$ diverges for non-interacting systems, i.e. for a vanishing spectral width, corresponding to infinite mean free path. Close to this limit $\eta$ can be expanded in a Laurent series (as realized for example analytically in ChPT and $\lambda\phi^4$ theory \cite{LangKaiserWeise2012}):
\begin{equation}
\label{LaurentExpEtaGamma}
  \eta[\Gamma]=\frac{A_{-1}}{\Gamma}+A_0+A_1\Gamma+A_2\Gamma^2+\ldots
\end{equation}
For small $\Gamma$, the combination $\eta\cdot\Gamma$ is just the residue $A_{-1}$. What does ``small'' mean in this context? In contrast to $\lambda\phi^4$ theory where $\Gamma\sim\lambda^2$, the NJL model is generically non-perturbative in its coupling, even though the scaling $G\sim 1/\Nc$ applies. The spectral width is therefore not expected to be sufficiently small in order to permit an expansion as in Eq.~\eqref{LaurentExpEtaGamma}.  Fig.~\ref{Fig5} shows results of the fully non-perturbative calculation of $\eta\cdot\Gamma$ as a function of the inverse width, conveniently written as $x=m_\pi/\Gamma$, at different $T$ and $\mu$ in comparison with the residue $A_{-1}$. As it can be seen from the figure, corrections to the leading term of the Laurent series \eqref{LaurentExpEtaGamma} are small for $x>1.5$ (demanding $10\%$ accuracy or better). From these considerations we conclude that a perturbative approach is justified only for spectral widths $\Gamma\ll\,m_\pi=140\;{\rm MeV}$.

The discussion of a perturbative treatment of $\eta[\Gamma(p)]$ is closely related to the resummation of ladder diagrams: if in the large-$\Nc$ limit the spectral width decreases, i.e. $\Gamma\sim 1/\Nc$ as suggested by hot-QCD calcuations where the coupling $\alpha_{\rm s}\sim 1/\Nc$ becomes small, then the perturbative regime is reached in this limit and the Laurent series expansion in \eqref{LaurentExpEtaGamma} can be restricted to its leading-order term. As seen from Eq.~\eqref{EtaSmallGamma}, for a constant but small spectral width $\Gamma$ the residue $A_{-1}$ can be identified with the remaining $\eps$-integral. While only this residue term of $\eta[\Gamma(p)]$ is relevant in this case, ladder diagrams now become sizable corrections and contribute also at leading order. Furthermore, the shear viscosity now scales as $\eta\sim\Nc^2$ and no longer linearly with $\Nc$ as Eqs.~\eqref{EtaLeadingNcAsTwoSpectralDensityPap} and \eqref{EtaLeadingNcAsSpectralWidth} do for $\Nc$-independent spectral function $\
rho$ and width $\Gamma$, respectively.

We conclude that ladder diagram resummations are necessary in the perturbative regime of $\eta[\Gamma(p)]$ in Eq.~\eqref{EtaLeadingNcAsSpectralWidth}, i.e. when the spectral width is small,  $\Gamma\ll m_\pi$. In the NJL model with its genuine non-perturbative structure, the physical spectral width is large and outside the perturbative regime. Ladder diagram resummations are subleading corrections, while the shear viscosity functional \eqref{EtaLeadingNcAsSpectralWidth} is valid also for large spectral width when including all orders of the Laurent series expansion \eqref{LaurentExpEtaGamma}.

\section[Effects of thermal quark masses and the ratio $\eta/\lowercase{s}$]{Effects of thermal quark masses and the ratio {\boldmath $\eta/\lowercase{s}$}} \label{Sec:ThermoMass}

The constituent quark mass has so far been treated as a constant. We now proceed to incorporate its explicit $T$ and $\mu$ dependence. One of the key features of the NJL model is the spontaneous breaking of chiral symmetry: ${\rm SU}(2)_{\rm L}\times{\rm SU}(2)_{\rm R}\to{\rm SU}(2)_{\rm V}$. In addition, chiral symmetry is explicitly broken by the non-zero current quark mass $m$. Solving the NJL gap equation results in a dynamically generated constituent quark mass \cite{Nambu1961tp,Nambu1961,VoglWeise1991}:
\begin{equation}
\begin{aligned}
\label{ThermalGapEquation}
  &M(T,\mu)=m-G\langle\bar{\psi}\psi\rangle =\,\\  
  &\!\!=m+\frac{2G\Nc M(T,\mu)}  {\pi^2}\int_0^\Lambda\dint p\,\frac{p^2}{E_p}\,\big[1-n_{\rm F}^+(E_p)-n_{\rm F}^-(E_p)\big],
\end{aligned}
\end{equation}
with $E_p(T,\mu)=\sqrt{p^2 + M^2(T,\mu)}$ and the Fermi-Dirac distribution for quarks and antiquarks
\begin{equation}
\label{DefFermiDistrMuT}
  n_{\rm F}^\pm(E)=n_{\rm F}(E\mp\mu)=\frac{1}{\e^{\beta (E\mp\mu)}+1}\;.
\end{equation}
In the vacuum, $(T,\mu)=(0,0)$, the mass is determined to be about one third of the nucleon mass, $M=325\;{\rm MeV}$, where the input parameters of the NJL Lagrangian are chosen as $G=10.08\;{\rm GeV^{-2}}$, $m=5.5\;{\rm MeV}$ and $\Lambda=650\;{\rm MeV}$. This parameter set produces physical (vacuum) values of the pion mass, $m_\pi=140\;{\rm MeV}$, the pion decay constant, $f_\pi=94\;{\rm MeV}$, and the chiral condensate $\langle\psib\psi\rangle=-(316.4\;{\rm MeV})^3$.

\begin{figure}[t!]
\begin{center}
  \subfigure[Parametric dependence on the dynamical constituent quark mass $M$ for a constant spectral width $\Gamma=100\;{\rm MeV}$]{\includegraphics[width=0.49\textwidth]{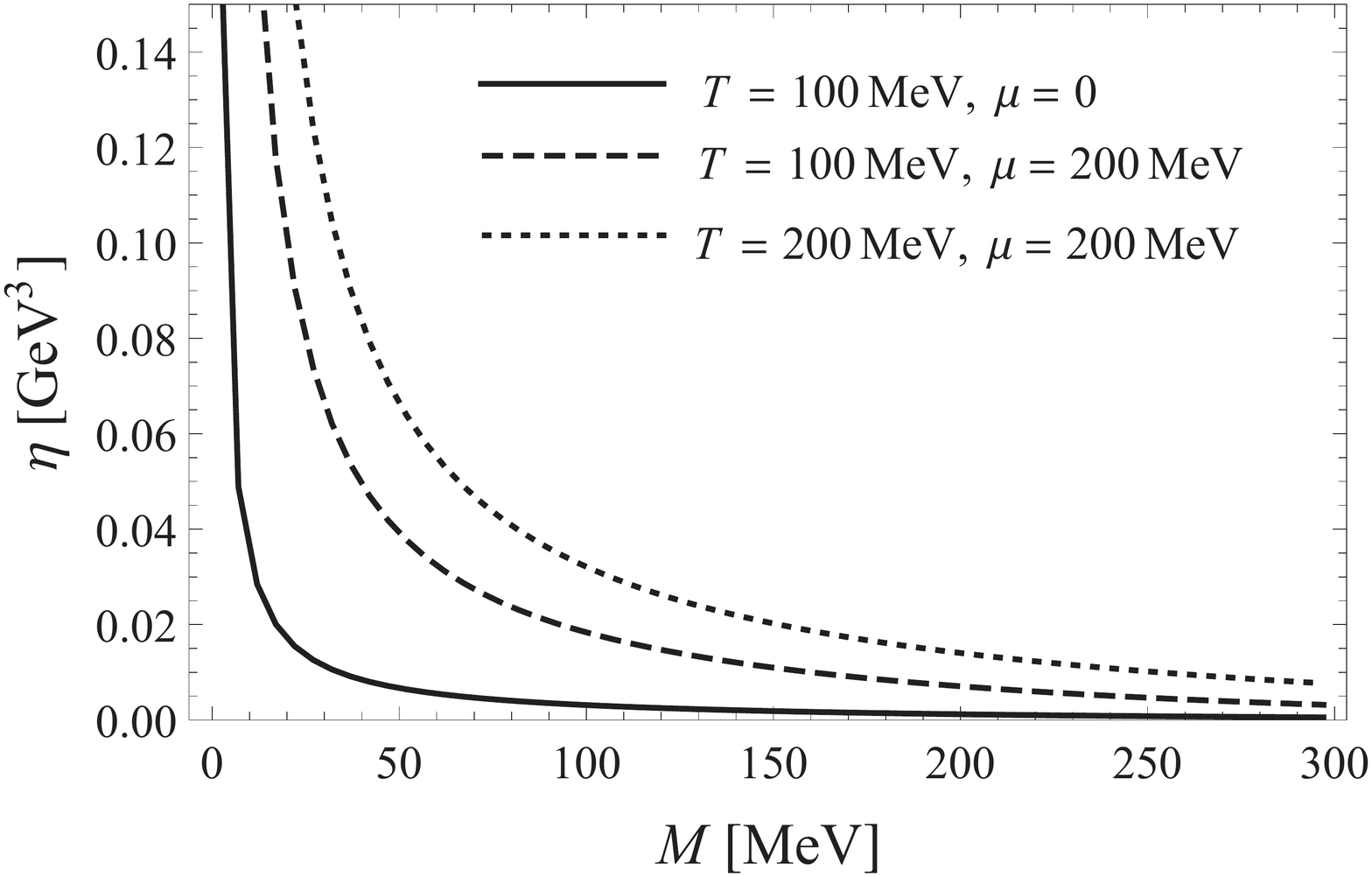}}
  \subfigure[Shear viscosity with thermal quark mass $M(T,\mu = 0)$]{\includegraphics[width=0.49\textwidth]{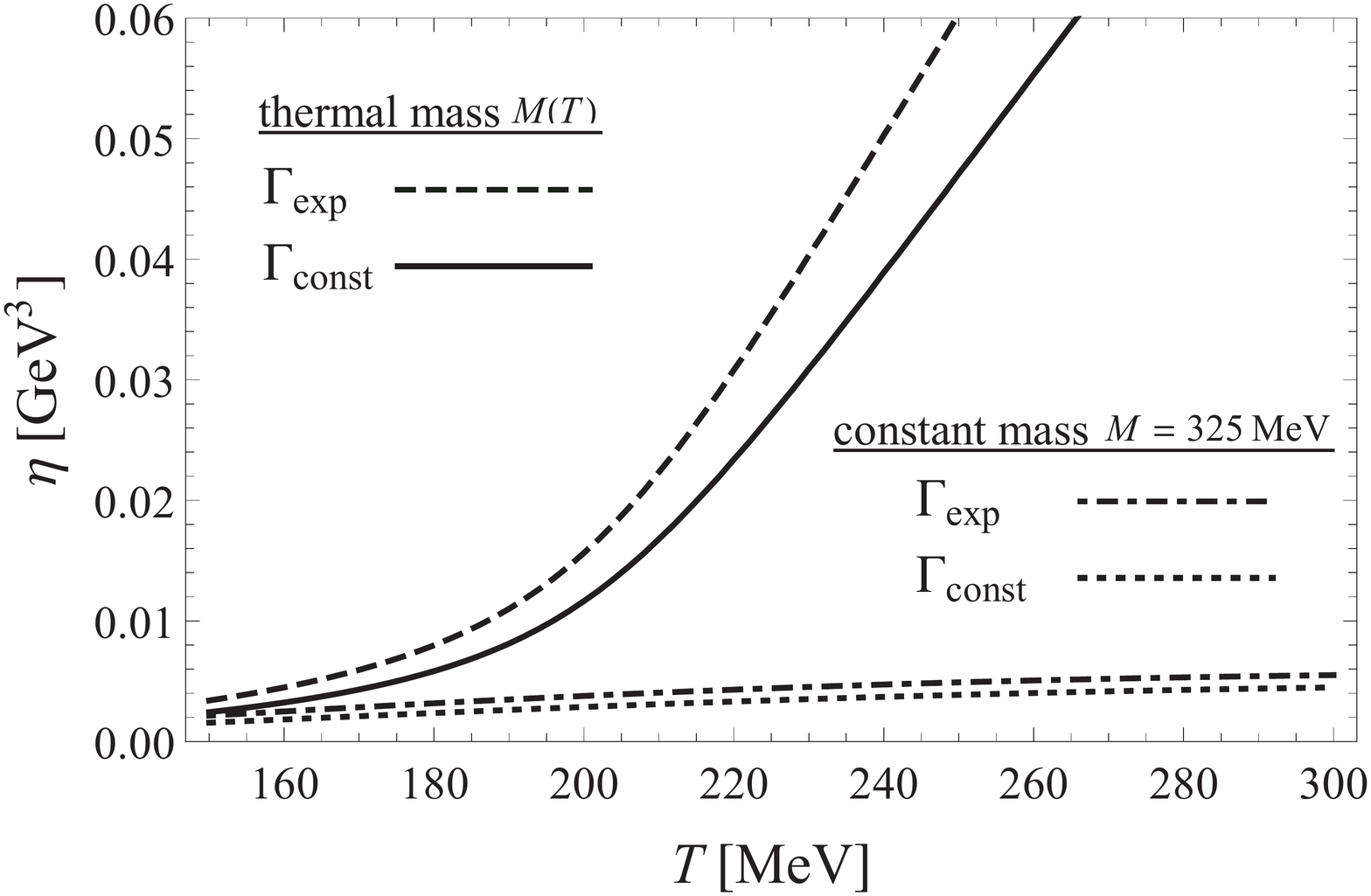}}
\caption{Dependence of the shear viscosity on the constituent quark mass: (a) parametric dependence on $M$ assuming a constant spectral width; (b) temperature dependence of the shear viscosity including the full thermal constituent quark masses $M(T,\mu)$ (upper lines, i.e. without dots) compared to results with a constant constituent quark mass $M=325\;{\rm MeV}$ (lower lines, i.e. with dots).}
\label{Fig6}
\end{center}
\end{figure}
Fig.~\ref{Fig6}(a) shows the shear viscosity $\eta$ for varying constituent quark mass $M$ treated as a parameter, assuming a constant spectral width $\Gamma_{\rm const}=100\;{\rm MeV}$. For $M\to 0$ the shear viscosity becomes divergent, again due to ``pinch poles'' appearing in Eq.~\eqref{EtaLeadingNcAsSpectralWidth} in this limit. In fact, the origin of this divergence is the same as for $\Gamma\to 0$, since $M$ and $\Gamma$ are formally (almost) interchangeable in the integrand of Eq.~\eqref{EtaLeadingNcAsSpectralWidth}. For large constituent quark masses, two effects occur: first, the maximizer $\eps^*(p)\sim M$ in Eq.~\eqref{EpsStarMaximizer} moves to larger values, and second, the integrand scales as $M^{-6}$. Both features result in a decreasing function $\eta(M)$.

\begin{figure}[t!]
\begin{center}
  \subfigure[Unphysical behavior of $\eta/s$ for a \textit{constant} constituent quark mass $M=100\;{\rm MeV}$, realized at $T\approx 200\;{\rm MeV}$]{\includegraphics[width=0.49\textwidth]{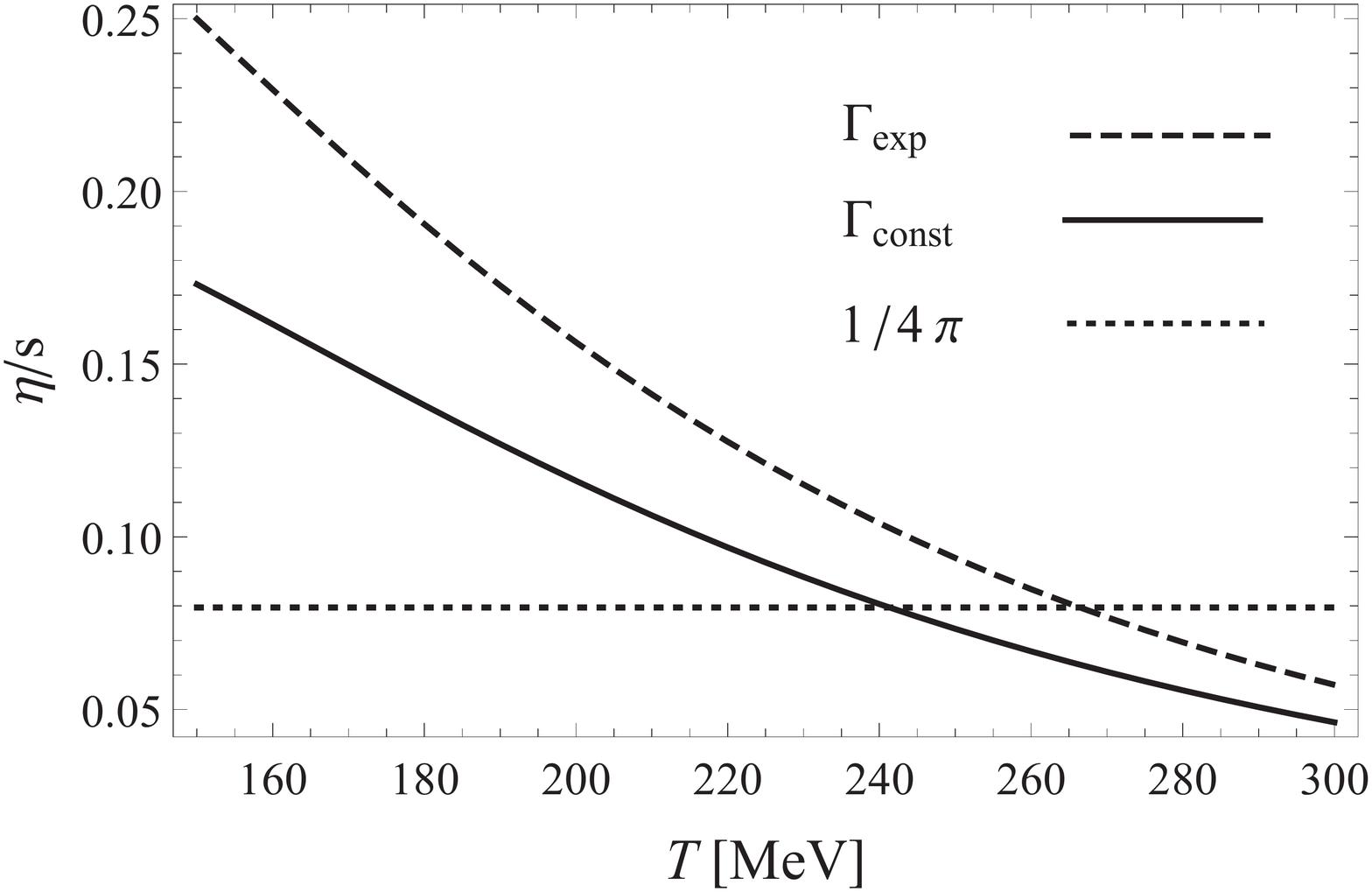}} \hspace{1cm}
  \subfigure[Thermal constituent quark mass $M(T,\mu = 0)$]{\includegraphics[width=0.49\textwidth]{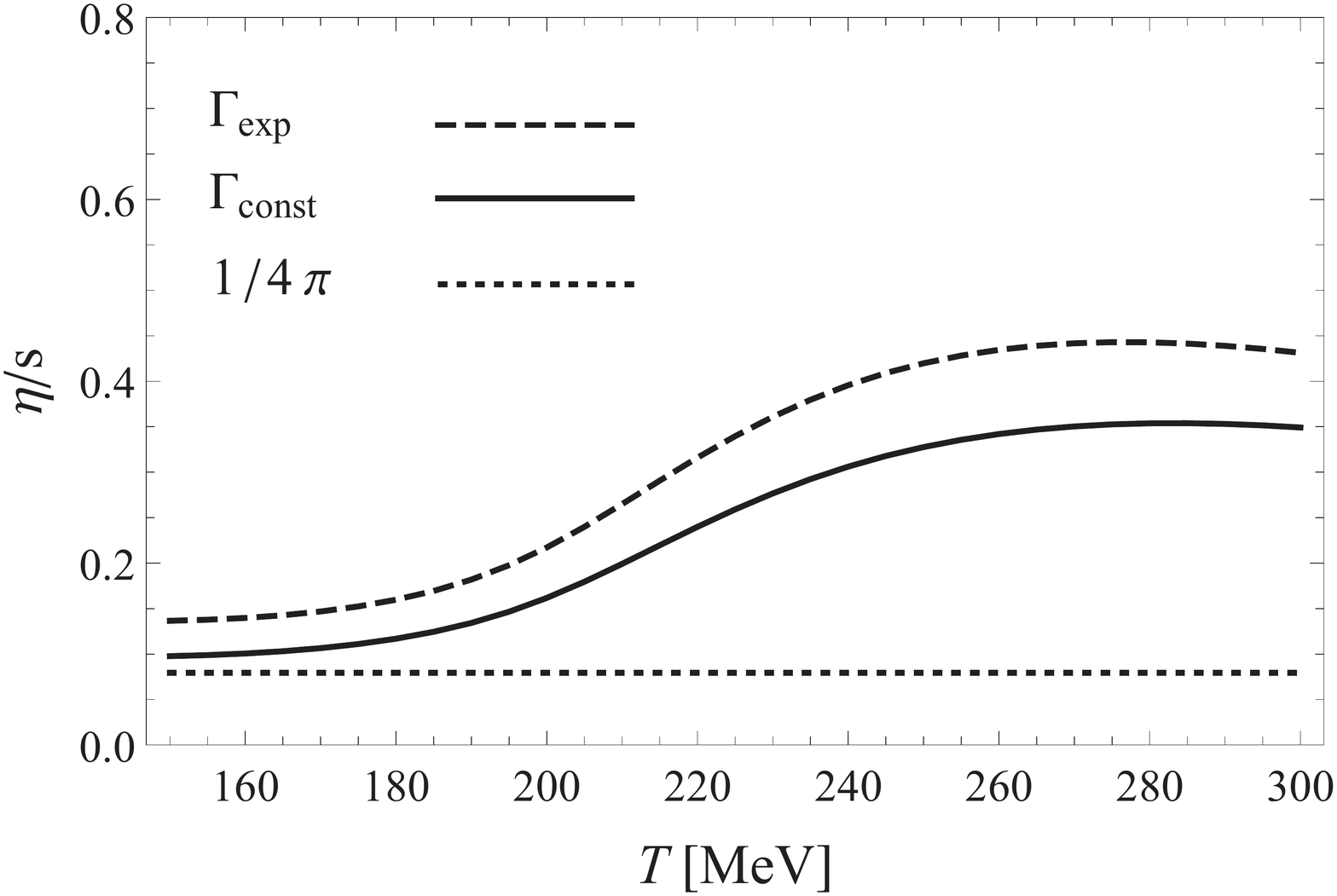}}
\caption{Ratio $\eta/s$ as function of temperature for a constant and exponentially damped spectral width at vanishing chemical potential. The entropy density $s$ is consistently derived within in the NJL model. Panel (a): no thermal effects, constant constituent quark mass $M=100\;{\rm MeV}$; panel (b): thermal constituent quark mass included. In both panels the horizontal dotted lines denote the AdS/CFT benchmark.}
\label{Fig7}
\end{center}
\end{figure}

Taking the full thermal dependence of the constituent quark mass into account has an essential influence on the shear viscosity, see Fig.~\ref{Fig6}(b): for small $T$, a constant mass $M=325\;{\rm MeV}$ approximates the thermal constituent quark mass. In contrast, at large $T$, with a melting chiral condensate, the dropping dynamical quark mass implies a strongly increasing shear viscosity, qualitatively
different from the case with constant quark mass. (We have chosen to compare thermal and non-thermal results for constant and exponential parameterizations of the spectral width. For $\Gamma_{\rm Lor}$ and $\Gamma_{\rm div}$ the results are qualitatively similar and therefore not shown.)

The shear viscosity itself is a dimensionful quantity. One usually compares $\eta$ to the entropy density $s$, in terms of the dimensionless ratio $\eta/s$. The corresponding ratio extracted from heavy-ion collisions is the smallest value of $\eta/s$ found so far in nature \cite{HeinzShenSong2012}. The entropy density of quark matter is given by \cite{Kapusta}:
\begin{equation}
\begin{aligned}
  s(T,\mu)=\frac{\Nc\Nf}{\pi^2}\int_0^\infty\dint p\,p^2\, &\left[\ln\left(1+\e^{-\beta E_p^+}\right) \right.\\
  & \left. \hspace{-4cm}+\ln\left(1+\e^{-\beta E_p^-}\right)+\frac{\beta E_p^+}{1+\e^{\beta E_p^+}}+\frac{\beta E_p^-}{1+\e^{\beta E_p^-}}\right],
\end{aligned}
\end{equation}
with $E_p^\pm=E_p\mp\mu$. The momentum integration ranges up to infinity.  It can be performed  without regularization. In the NJL model the thermal constituent quark mass, for momenta above the cutoff scale $\Lambda=650\;{\rm MeV}$, reduces to the bare current quark mass, $M\rightarrow m=5.5\;{\rm MeV}$. At low temperatures, $T\lesssim 150$ MeV, confinement implies that quarks are not the relevant physical degrees of freedom any more and the NJL model cannot be expected to give a realistic description of transport properties. 

The evaluation of the ratio $\eta/s$ is shown in Fig.~\ref{Fig7} for $\Nf=2$ and $\Nc=3$, for the cases of a constant and an exponentially damped spectral width. The comparison between the panels (a) and (b) clearly demonstrates that implementing thermal masses is crucial in order to avoid an unphysical, continuously decreasing ratio $\eta/s(T)$. From experimental data \cite{Lacey07} and lattice calculations \cite{Nakamura05,Meyer07} it is in fact suggested that this ratio increases for $T> 200\;{\rm MeV}$. We also compare to the benchmark $\eta/s=1/4\pi$ from AdS/CFT correspondence \cite{Maldacena99,Kovtun05}. This value is known not to be a universal bound; it can be undershot in some field theories \cite{Cohen07,Rebhan12,Mamo12}. This is also found in Fig.~\ref{Fig7}(a), additionally to the unphysical evolution of $\eta/s$ with increasing temperature. A constant quark mass $M=100\;{\rm MeV}$ has been chosen there for convenience. Its vacuum value, $M=325\;{\rm MeV}$, would reduce the scale of the $\eta/s$ 
ratio even more, compare Fig.~\ref{Fig6}(a). However, taking the thermal constituent quark masses into account leads to an increasing function $\eta/s(T)$, see Fig.~\ref{Fig7}(b). Despite the fact that $\eta$ itself rapidly increases at high $T$, the ratio $\eta/s$ flattens in that region. This flattening is expected to be shifted to higher temperatures for more realistic forms of the width $\Gamma(p)$ such the Lorentzian. In the considered temperature range $\eta/s$ stays above the AdS/CFT benchmark for all parameterizations of the width $\Gamma$. 

\section{Spectral width at one-loop level} \label{Sec:GammaNJL}
So far we have discussed the impact of the shape of the spectral width on the shear viscosity of quark matter, its strong sensitivity to the NJL cutoff and to the thermal constituent quark masses. In this section an explicit calculation is performed including one-loop mesonic contributions at next-to-leading order in the large-$\Nc$ expansion \cite{BuballaMuellerWambach2010}. The leading-order gap equation \eqref{ThermalGapEquation} is modified by the mesonic insertions
\begin{equation}
  \label{NJLMesonInsertion}
\Sigma^{\rm S/P}_\beta(\vec{p},\nu_n) = \includegraphics[width=0.15\textwidth]{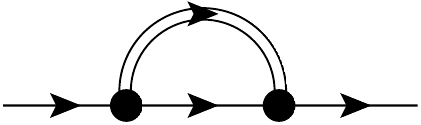}\;,
\end{equation}
where $\nu_n=(2n+1)\pi T-\i\mu$ are the Matsubara frequencies for quarks.  The corresponding expression for antiquarks is easily obtained  by inserting $\bar{\nu}_n=\nu_n^*$. The spectral width is extracted from the imaginary part of this self-energy\footnote{Note that in this definition as well as in previous expressions  the spectral width $\Gamma$ corresponds to half the FWHM (full width at half maximum) of the resulting quark spectral function, $\pi\rho(p)=-\Im\,G^{\rm R}(p)$, where the retarded quark propagator has been defined in Eq.~\eqref{KuboQuasiPartApprQuarkProp}.}:
\begin{equation}
  \Gamma^{\rm S/P}_{\rm q}(p_0,\vec{p})=-\lim_{\eps\to 0}\,{\rm Im}\,\Sigma^{\rm S/P}_\beta(\vec{p},-\i p_0+\eps)\;.
\end{equation}
It is of next-to-leading order using the mesonic modes generated by the Bethe-Salpeter equation (BSE) in the pertinent quark-antiquark channels (e.g. \cite{HatsudaKunihiro87, KLVW1990, BuballaMuellerWambach2010}). Whereas the gap equation \eqref{ThermalGapEquation} is of leading order, $\mathcal{O}(1)$ in $N_c$, the BSE is of order $\mathcal{O}(1/\Nc)$. In the two-flavor case the mesonic loop involves the three pions and the sigma mode which contribute with positive and negative signs, respectively, to the spectral width $\Gamma(p)$. This ``antiscreening'' by the sigma mode \cite{QuackKlevansky94} cancels roughly one of the three pionic contributions to the spectral width, as seen in Fig.~\ref{Fig9}a. The effective spectral width (for $N_f = 2$) reads:
\begin{equation}
\label{QuarkSpectralWidthOneLoopAllMesons}
 \Gamma(p)=3\Gamma_{\rm q}^{\rm P}(p)+\Gamma_{\rm q}^{\rm S}(p)\;,
\end{equation}
with the scalar and pseudoscalar contributions, $\widetilde{N}^{\rm S}=-2$ and $\widetilde{N}^{\rm P}=2$, respectively:
\begin{figure}[t!]
\begin{center}
  \subfigure[Landau damping]{\includegraphics[width=0.2\textwidth]{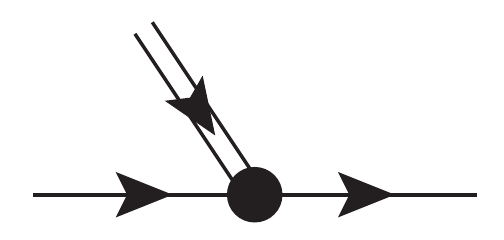}}
  \subfigure[Recombination process]{\includegraphics[width=0.2\textwidth]{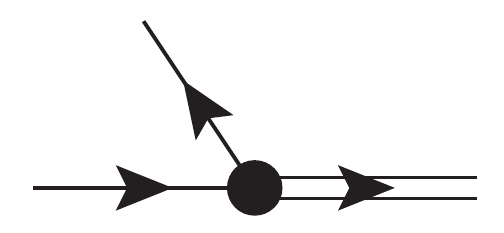}}
  \caption{Dissipative processes contributing to the shear viscosity at one-loop level, i.e. next-to-leading order in $1/\Nc$}
  \label{Fig8}
\end{center}
\end{figure}
\begin{equation}
\label{QuarkSpectralWidthOneLoop}
  \Gamma^{\rm S/P}_{\rm q}(p)=\frac{M\gMq^2 \widetilde{N}^{\rm S/P}}{4\pi |\vec{p}|}\int_{E_{\rm min}}^{E_{\rm max}} \dint E_f \left[n_{\rm B}(E_b)+n_{\rm F}^-(E_f)\right],
\end{equation}
involving the Fermi distribution for antiquarks, $n_{\rm F}^{-}$ in Eq.~\eqref{DefFermiDistrMuT}, and the Bose distribution for mesons at zero (quark-)chemical potential:
\begin{equation}
\label{BoseDistribution}
  n_{\rm B}(E)=\frac{1}{\e^{\beta E}-1}\;.
\end{equation}
Energy-momentum conservation implies $E_b=E_f+p_0$ in Eq.~\eqref{QuarkSpectralWidthOneLoop} and leads to the restricted range of integration:
\begin{equation}
\begin{aligned}
\label{QuarkSpectralWidthEminEmax}
  E_{{\rm min},{\rm max}}(p)&=\frac{1}{2M^2}\left[\left(m_{\rm M}^2-2M^2\right)\sqrt{M^2+p^2}\right. \\
  &\hspace{2cm}\left.\mp\, p\,m_{\rm M}\sqrt{m_{\rm M}^2-4M^2}\right].
\end{aligned}
\end{equation}
Here $m_{\rm M}$ denotes the $T$- and $\mu$-dependent (pseuodoscalar or scalar) meson mass and $M(T,\mu)$ is the dynamical quark mass in the thermal medium, c.f. Eq.~\eqref{ThermalGapEquation}. The quark-meson vertex is of Yukawa type with a (thermal) coupling constant $g_{\rm Mqq}(T,\mu)$ arising from the renormalized meson propagator $D_{\rm M}(p_0,\vec{p})$ at $(p_0,\vec{p})=(m_{\rm M},\vec{0})$:
\begin{equation}
  g_{\rm Mqq}=\left({\rm Res}\, D_{\rm M}\right)^{-1/2}\;.
\end{equation}
The details of the derivation of Eqs.~\eqref{QuarkSpectralWidthOneLoop} and \eqref{QuarkSpectralWidthEminEmax} are presented in the Appendix \ref{Sec:App}. The spectral width is a function of the energy $p_0$ and the momentum $p=|\vec{p}|$ of the quark propagating in the (isotropic) thermal medium. In the on-shell case, $p_0=\sqrt{p^2+M^2}$, the width $\Gamma$ becomes a function of $p$ only.

Taking the imaginary part of the mesonic insertion means cutting the loop diagram and forcing it on-shell. Two different dissipative mechanisms contribute to the width, as sketched in Fig.~\ref{Fig8}. The term involving the Bose distribution describes Landau damping, a process also known from high-$T$ QCD calculations \cite{Boyanovsky1998}, and recently applied to the calculation of the shear viscosity from kinetic theory \cite{Ghosh2013}. In the NJL model Landau damping arises from the scattering of quarks on the mesonic collective modes in the thermal medium. This scattering process dissipates energy from the quark sector and contributes to the shear viscosity. The second mechanism is a recombination process: a collective mesonic mode is created by quark-antiquark rescattering. This is described by the term involving the Fermi distribution of thermal antiquarks in Eq.~\eqref{QuarkSpectralWidthOneLoop}.

\begin{figure}[t!]
\begin{center}
  \subfigure[Contributions from single pion and sigma modes]{\includegraphics[width=0.49\textwidth]{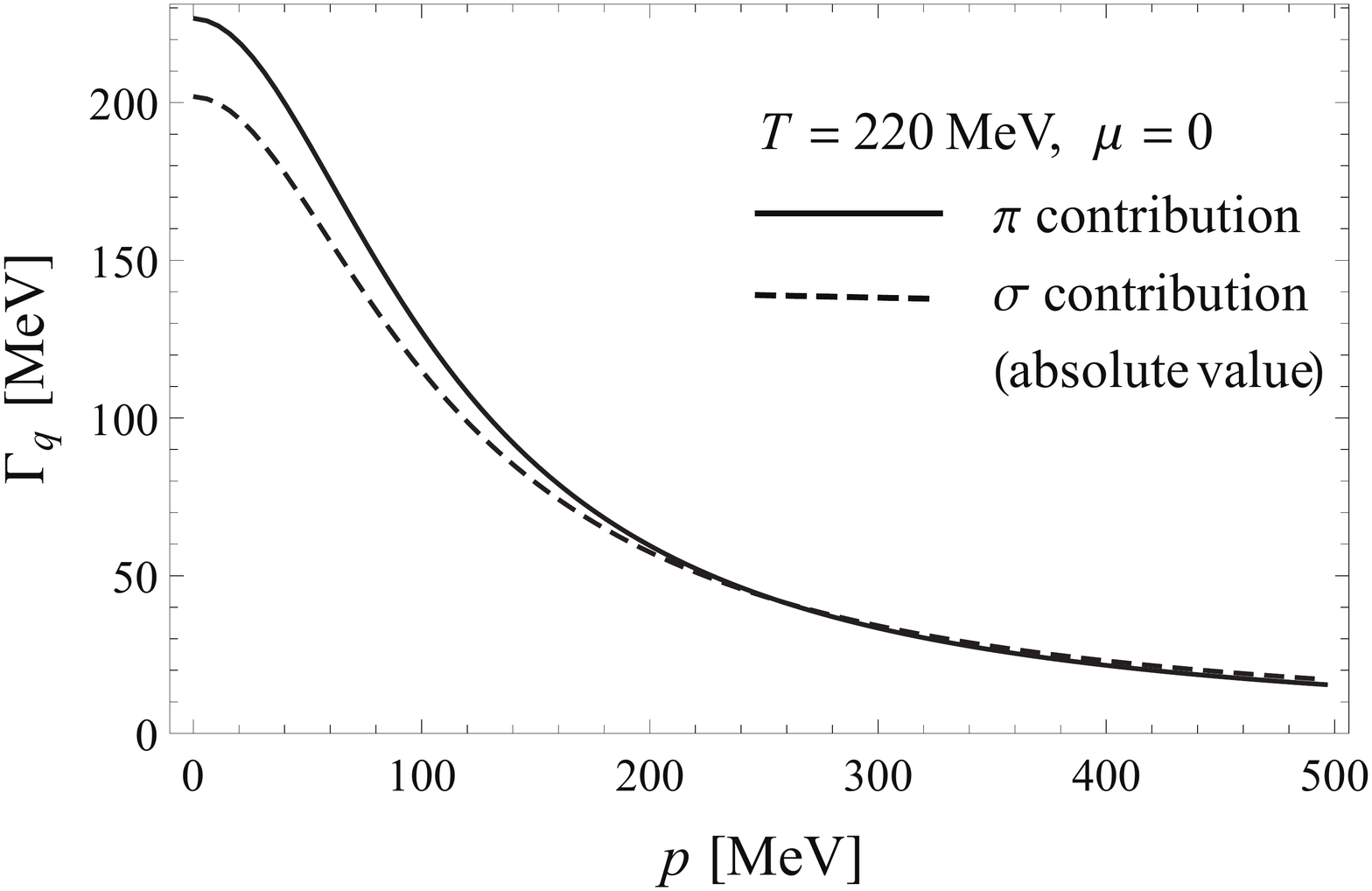}}
  \subfigure[Spectral width summed over all pion and sigma modes for different temperatures.]{\includegraphics[width=0.49\textwidth]{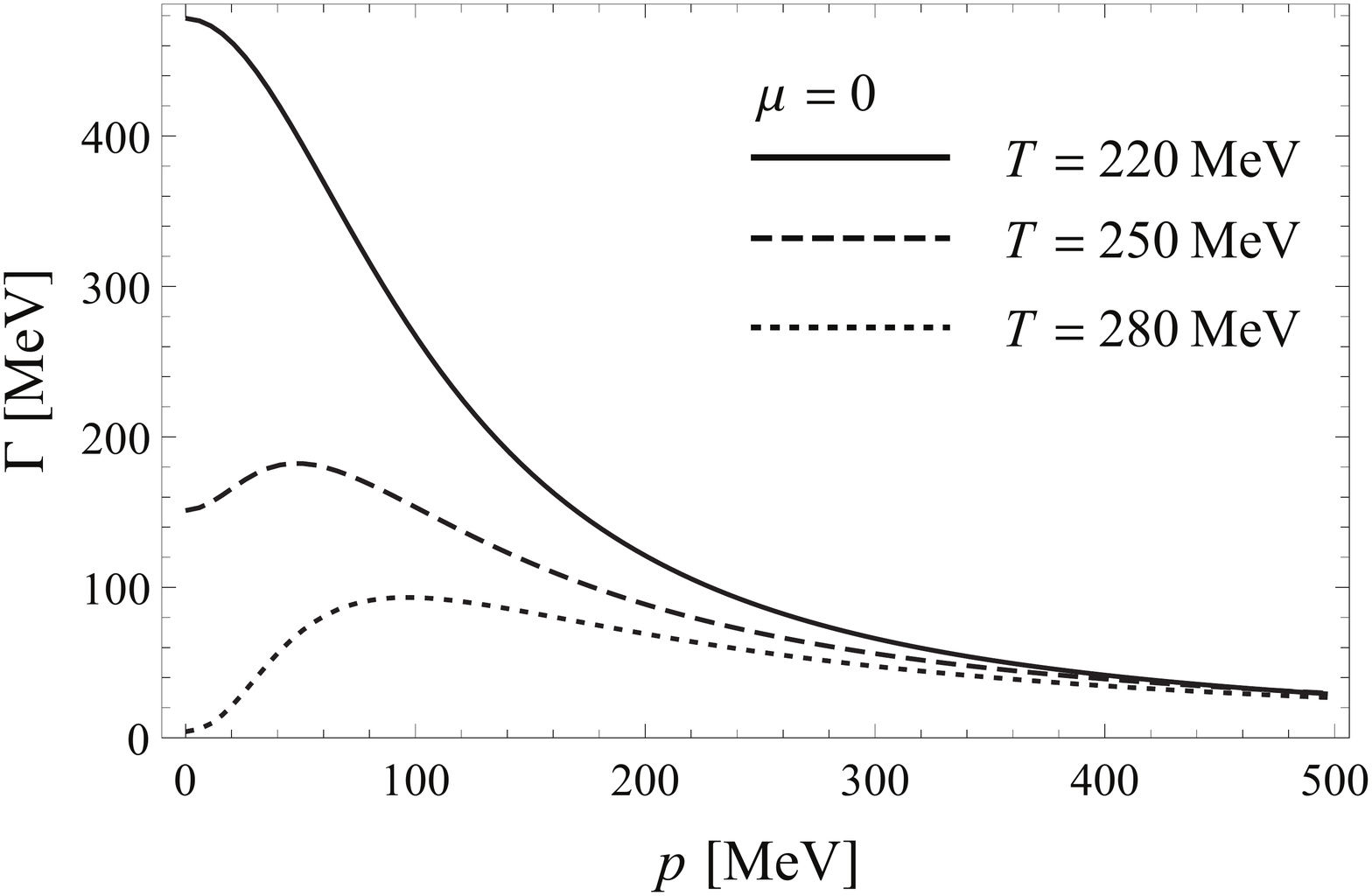}}
  \caption{Spectral width from one-loop mesonic contributions to the quark self-energy at vanishing chemical potential}
  \label{Fig9}
\end{center}
\end{figure}

Results for the calculated on-shell spectral width including all mesons, $\Gamma(p)$ in Eq.~\eqref{QuarkSpectralWidthOneLoopAllMesons}, are shown in Fig.~\ref{Fig9}b. Due to the ``antiscreening'' caused by the sigma mode, the total $\Gamma(p)$ is rougly twice the single pion contribution, c.f. Fig.~\ref{Fig9}a. For small quark momenta the spectral width can become quite large. At $p\approx 200\;{\rm MeV}$ it is of the order of $100\;{\rm MeV}$, roughly as large as the dynamical (constituent) quark mass. We recall from Fig.~\ref{Fig5} that the shear viscosity $\eta[\Gamma]$ can be treated perturbatively only if $\Gamma\ll m_\pi$. Hence for the one-loop spectral width calculated in the NJL model, the perturbative regime is not reached, given the large values of $\Gamma(p)$. With rising temperature the spectral width decreases leading to an increasing shear viscosity $\eta(T)$, as explored parametrically in Fig.~\ref{Fig1}. One important reason for this behavior is the range of integration, $E_f(p)\in[E_{\rm 
min}(p),E_{\rm max}(p)]$, which has its support for $m_\pi>2M$, the kinematic threshold condition for a pion to decay on-shell into two constituent quarks. As the temperature increases this range of integration is shifted to higher values of $E_f$ and $p$, but these are exponentially suppressed by the Bose and Fermi distributions in Eq.~\eqref{QuarkSpectralWidthOneLoop}. For more details, see also Fig.~\ref{Fig11} in the Appendix.

\begin{figure}[t!]
\begin{center}
  \includegraphics[width=0.49\textwidth]{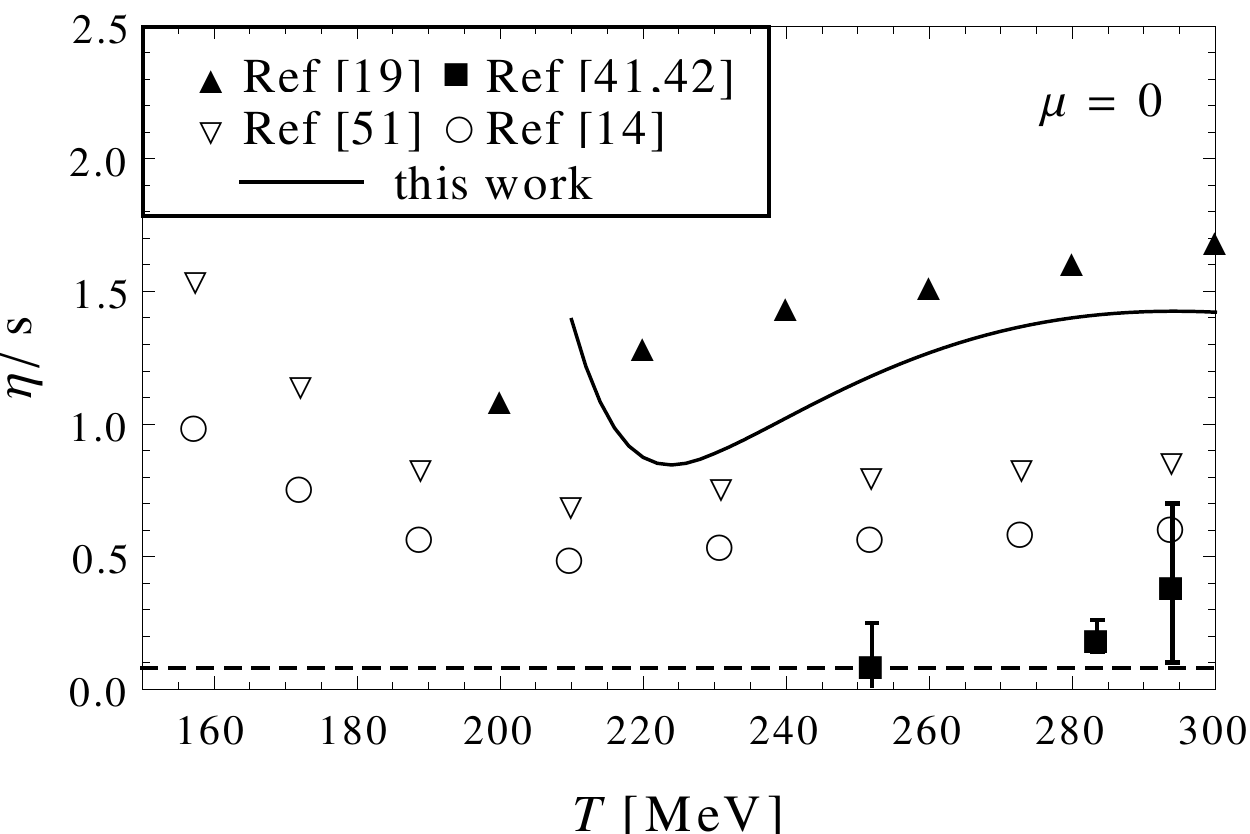}
  \caption{Shear viscosity per entropy density at zero chemical potential from the two-flavor NJL model at one-loop level in a large-$\Nc$ expansion (solid line). Shown for comparison are results using the kinetic approach (open circles and triangles), those of a related calculation using the Kubo formalism (solid triangles) and pure-gauge lattice data (solid squares). The dashed horizontal line is the AdS/CFT benchmark $1/4\pi$.}
  \label{Fig10}
\end{center}
\end{figure}

Finally the resulting shear viscosity $\eta(T,\mu = 0)$ (its ratio $\eta/s$) is shown in Fig.~\ref{Fig10}. Only temperatures large enough to satisfy $m_\pi>2M$ give rise to a finite shear viscosity. Therefore the one-loop results are restricted to $T\gtrsim 210\;{\rm MeV}$. We compare our results to those from a kinetic approach using the two-favor NJL model \cite{SasakiRedlich2010} (open circles) and \cite{Ghosh2013} (open triangles). Our values based on the Kubo formalism are larger but feature the same order of magnitude and the same qualitative behavior. These findings are consistent with the general observation that the kinetic approach seems to underestimate the shear viscosity as pointed out in Refs. \cite{PlumariEtAl2012,PlumariEtAl2013}. Their calculations have also been performed using the Kubo formalism (c.f. Fig.~8 in \cite{PlumariEtAl2012}) and lie slightly above our result for $\eta/s$ as shown in Fig.~\ref{Fig10} (filled triangles). A flattening of $\eta/s$ is observed at higher temperatures 
around $T\approx 300\;{\rm MeV}$, indicating that the shear viscosity scales roughly as $T^3$ in this range. The same behavior has been found in our preceding parameter study, c.f. Fig.~\ref{Fig6}. All results for $\eta/s$, both from kinetic and Kubo approaches, exceed those from  pure-gauge lattice QCD \cite{Nakamura05,Meyer07} that are close to the AdS/CFT bound as shown by the solid squares in Fig.~\ref{Fig10}.

\section{Summary and Conclusions}
In the present work we have used the Kubo formalism to derive a general functional $\eta[\Gamma(p)]$ for a class of fermionic theories, in particular the two-flavor NJL model with scalar and pseudoscalar interactions. At leading order in $1/\Nc$ the retarded correlation function $\Pi^{\rm R}(\omega)$ reduces to a single generic diagram at $\mathcal{O}(\Nc^1)$. We have found that it is not necessary to work in the chiral limit to obtain this result derived previously in \cite{Fukutome2006,Fukutome2008Nucl,Fukutome2008Prog} using stronger assumptions.

The detailed study of the functional $\eta[\Gamma(p)]$ leads to a convergence criterion, Eq.~\eqref{CriterionEtaGamma}, to be fulfilled by the spectral width $\Gamma(p)$ in order for the shear viscosity $\eta$ to be finite without regularization. Four different schematic parameterizations have been chosen for $\Gamma(p)$ and a numerical approximation scheme suitable for arbitrary, momentum-dependent spectral widths has been introduced. The results for $\eta$ show a strong cutoff dependence: restricting the momentum integration to the typical NJL cutoff, $p<\Lambda=650\;{\rm MeV}$,  changes the shear viscosity drastically as shown in Fig.~\ref{Fig3}. Such a cutoff places $\eta$ at physically meaningful values by excluding up to $90\%$ of the numerically accessible high-momentum region, as demonstrated in Fig.~\ref{Fig4}.

An exploration has been performed determining the range of $\Gamma(p)$ for which a perturbative treatment of the (generally non-perturbative) shear viscosity is adequate. The functional $\eta[\Gamma(p)]$ as given in Eq.~\eqref{EtaLeadingNcAsSpectralWidth} is valid in particular for a large spectral width as it is expected in the NJL model. For a small spectral width, $\Gamma\ll m_\pi$, a Laurent series expansion of $\eta$ with restriction to the leading term $\eta \sim 1/\Gamma$ is justified. In this limit, subleading 
effects from resummations of ladder diagrams are expected to become non-negligible as known from 
studies in bosonic field theories \cite{JeonSkeleton1995,HidakaKunihiro2011}. On the other hand, for 
the NJL model with its comparatively large spectral width, it is consistent to omit ladder diagram resummations while taking all orders of the Laurent expansion in the width $\Gamma(p)$ into account.

Including thermal quasiparticle masses (the temperature and density dependence of the dynamically generated constituent quark masses in NJL-like models) is of crucial importance.  As experimental data indicate, the ratio $\eta/s$ is expected to increase for high temperatures $T > T_{\rm c}$ where $T_{\rm c}\approx 180-200\;{\rm MeV}$ is the typical temperature interval of the chiral crossover. A constant
quark mass would instead lead to a continuously decreasing ratio $\eta/s$ with increasing temperature, as seen in Fig.~\ref{Fig7}. In comparison with other approaches to the shear viscosity using NJL-type models, such as the Boltzmann equation in relaxation time approximation, one can find different results: an increasing ratio $\eta/s(T)$ for two flavors and restricting to scalar and pseudoscalar interactions \cite{SasakiRedlich2010}, but also a decreasing behavior \cite{MartyFrankfurt2013,HidakaKunihiro2011NJL}.

From AdS/CFT correspondence, the scaling of the ratio $\eta/s$ is expected to be of $\mathcal{O}(\Nc^0)$ in the large-$\Nc$ limit. With the entropy density of quarks and the functional $\eta[\Gamma(p)]\sim\Nc$ in Eq.~\eqref{EtaLeadingNcAsSpectralWidth}, we find indeed $\eta/s\sim \mathcal{O}(1)$ in the NJL model. 
Its applicability is, however, limited to the temperature range $T_{\rm c} \lesssim T < \Lambda$. At much higher temperatures gluonic degrees of freedom are dominant and contribute to the entropy density as $s_{\rm gluon}\sim\Nc^2$ due to their adjoint representation instead of the fundamental representation for fermions. At the same time the spectral width is expected to became small ($\Gamma\to 0$) and scales as $1/N_c$, leading to ``pinch poles'' and a scaling of the shear viscosity as $\eta \sim N_c^2$, as seen from Eq.~\eqref{EtaSmallGamma}. Consequently the ratio still scales as $\eta/s\sim\mathcal{O}(1)$ at large $T$ in the limit $\Nc\to\infty$. 

From the systematic parameter study the conclusion can be drawn that the momentum dependence of the spectral width $\Gamma(p)$ determines primarily the overall scale of the ratio $\eta/s$. Its actual behavior as a function of temperature is largely governed by the thermal properties of the quark propagator. A consistent one-loop NJL model calculation of the quark self-energy in the thermal medium has been performed at next-to-leading order in the large-$\Nc$ expansion. At this order the gap equation for the thermal constituent quark mass is corrected by the mesonic insertions \eqref{NJLMesonInsertion}. Two generic dissipative contributions to the shear viscosity emerge: Landau damping and a quark-antiquark recombination process, resulting in the spectral width $\Gamma(p)$ given in Eq.\,\eqref{QuarkSpectralWidthOneLoop}. The restricted range of the energy integration leads to a decreasing function $\Gamma(T)$ as the temperature increases, which implies an increasing shear viscosity for high temperatures as 
seen in Fig.~\ref{Fig10}.

Additional contributions to $\Gamma$ from quark-quark scattering via exchange of mesonic quark-antiquark modes still have to be included. Such processes appear at order $\Nc^{-1}$ in the large-$\Nc$ expansion and are calculated in ongoing work. Results will be reported elsewhere.

\section*{Acknowledgments}
This work is partially supported by BMBF and by the DFG Cluster of Excellence ``Origin and Structure of the Universe''. R. L. thanks the ECT* for kind hospitality. Valuable discussions with Y. Hidaka and T. Hatsuda are acknowledged. Thanks go to N. Kaiser for useful advices and reading of the manuscript. R.L. has been supported by the TUM Graduate School (TUM-GS), by the RIKEN IPA and the RIKEN iTHES projects.

\appendix
\section{Derivation of the NJL spectral width} \label{Sec:App}
In this appendix we focus on the analytical treatment of the mesonic insertion $\Sigma_\beta^{\rm S/P}$ in Eq.~\eqref{NJLMesonInsertion} and the extraction of the spectral width $\Gamma(p)$ as given in Eqs.~\eqref{QuarkSpectralWidthOneLoopAllMesons} and \eqref{QuarkSpectralWidthOneLoop}. The Matsubara frequencies for a thermal quark are $\nu_n=(2n+1)\pi T-\i\mu$, whereas they read for an antiquark: $\bar{\nu}_n=(2n+1)\pi T+\i\mu=\nu_n^*$. The diagram for the mesonic insertion represents both pions (pseudoscalar channel) and the sigma meson (scalar channel) contributing the quark spectral width:
\begin{equation}
\begin{aligned}
  \Sigma^{\rm S/P}_\beta(\vec{p},\nu_n)\; &=-4M g_{\rm Mqq}^2 \widetilde{N}^{\rm S/P} \\
  & \hspace{-1.5cm}\times\ThermalInt{m}{q}\,\frac{1}{\nu_m^2+E_f^2}\,\,\frac{1}{(\nu_m-\nu_n)^2+E_b^2}\;,
\end{aligned}
\end{equation}

with $E_f^2=\vec{q}^2+M^2$ and $E_b^2=(\vec{q}-\vec{p})^2+m_{\rm M}^2$. We have introduced $\widetilde{N}^{\rm P}=2$ and $\widetilde{N}^{\rm S}=-\Nf=-2$ reflecting the opposite parities of pion and sigma boson, leading to screening and antiscreening of the quark mass, respectively. Carrying out the Matsubara sum introduces combinations of Fermi and Bose distribution functions, Eq.~\eqref{DefFermiDistrMuT} and \eqref{BoseDistribution}, respectively:
\begin{equation}
\begin{aligned}
  Z_1 &=1+n_{\rm B}(E_b)-\frac 12\left(n_{\rm F}^+(E_f)+n_{\rm F}^-(E_f)\right), \\
  Z_2 &= n_{\rm B}(E_b)+\frac 12\left(n_{\rm F}^+(E_f)+n_{\rm F}^-(E_f)\right), \\ 
  Z_3 &=n_{\rm F}^+(E_f)-n_{\rm F}^-(E_f)\;.
\end{aligned}
\end{equation}
The thermal self-energy of the quark is given by:
\begin{widetext}
\begin{equation}
\label{WidthFirstItThermalSelfEnergy}
  \Sigma^{\rm S/P}_\beta\!=\!-4Mg_{\rm Mqq}^2\widetilde{N}^{\rm S/P}\!\!\int\frac{\dint^3q}{(2\pi)^3}\left[\frac{1}{2E_bE_f}\left[\frac{(E_f+E_b)Z_1}{(E_f+E_b)^2+\nu_n^2}+\frac{(E_f-E_b)Z_2}{(E_f-E_b)^2+\nu_n^2}\right]+\frac{\i\nu_n Z_3}{[(E_f+E_b)^2+\nu_n^2][(E_f-E_b)^2+\nu_n^2]} \right].
\end{equation}
\end{widetext}
The analytical continuation of the thermal self-energy to the retarded self-energy in Minkowskian spacetime,
\begin{equation}
  \Sigma^{\rm S/P}_{\rm R}(p_0,\vec{p}):=\Sigma^{\rm S/P}_\beta(\vec{p},-\i p_0+\eps)\;,
\end{equation}
leads to the spectral width
\begin{equation}
\label{IterationOneSpectralWidthPart1}
  \Gamma^{\rm S/P}_{\rm q}(p_0,\vec{p}) :=-\lim_{\eps\to 0}\Im\Sigma^{\rm S/P}_{\rm R}(p_0,\vec{p})\;.
\end{equation}
The non-vanishing spectral width is induced by the pole structure of the propagators:
\begin{equation}
\begin{aligned}
\label{DeltaFctArrImPart}
  \lim_{\eps\to 0}\Im\left.\frac{Z}{x^2+\nu_n^2}\right|_{\nu_n\mapsto -\i p_0+\eps} &= Z\pi\delta(x^2-(p_0)^2) \\ & \hspace{-1.5cm}=\frac{\pi Z}{2p_0}\left[\delta(x-p_0)+\delta(x+p_0)\right].
\end{aligned}
\end{equation}
This means for the $Z_1$ term: $E_f+E_b\pm p_0=0$, where only the minus sign can be realized. For the $Z_2$ term, $E_f-E_b\pm p_0=0$, both signs can be realized for the time being. We will see that only the plus sign contributes to the (on-shell) spectral width, so there is just one contribution from $Z_2$. The $Z_3$ term is considered later.

From the identity in Eq.~\eqref{DeltaFctArrImPart} the following structure is found for the $Z_1$ and $Z_2$ terms:
\begin{equation}
\label{CalcZ1Z2PhaseSpace}
\begin{aligned}
  &\int\frac{\dint^3q}{(2\pi)^3}\frac{\pi Z}{2p_0}\frac{1}{2E_bE_f}\,\delta(E_b-(*))= \\
  &=\int\frac{\dint^3q}{(2\pi)^3}\frac{\pi Z}{2p_0E_f}\,\delta(E_b^2-(*)^2)= \\
\end{aligned}
\end{equation}
\begin{equation*}
\begin{aligned}
  &=2\pi\int_{-1}^1\dint\xi\int_0^\infty \frac{\dint q\, q^2}{(2\pi)^3}\frac{\pi Z}{2p_0E_f }\delta(E_b^2(\xi)-(*)^2)=\\
  &=2\pi\int_{M}^\infty\frac{\dint E_f}{(2\pi)^3}\frac{\pi Z}{4p_0p}\,\Theta(1-\xi^2)\;,
\end{aligned}
\end{equation*}
with $\xi =\cos\theta$. In order to carry out this integral over the delta function we have used
\begin{equation}
  E_b^2 =m_{\rm M}^2+(\vec{p}-\vec{q})^2=m_{\rm M}^2+p^2+q^2-2pq\xi\;,
\end{equation}
from which follows
\begin{equation}
  \left|\frac{\partial E_b^2}{\partial\xi}\right|=2pq\;.
\end{equation}
In addition the integral over momentum has been translated into an energy integral using $q\,\dint q=E_f\,\dint E_f$. The ill-conditioned $\Theta$ term can be removed by the following consideration: The condition $|\xi|\leq 1$ is equivalent to
\begin{equation}
\begin{aligned}
\label{DefCapitalFfunctionPhaseSpace}
  -1\leq & \frac{E_b^2-m_{\rm M}^2-p^2-q^2}{2pq}\leq 1 \\
  & \Leftrightarrow F(E_f,p)\geq 0\;,
\end{aligned}
\end{equation}
where we have defined
\begin{equation}
F(E_f,p):=4p^2(E_f^2-M^2)-\left[E_b^2-m_{\rm M}^2-p^2+M^2-E_f^2\right]^2.
\end{equation}
\begin{figure}[t!]
\begin{center}
  \includegraphics[width=0.49\textwidth]{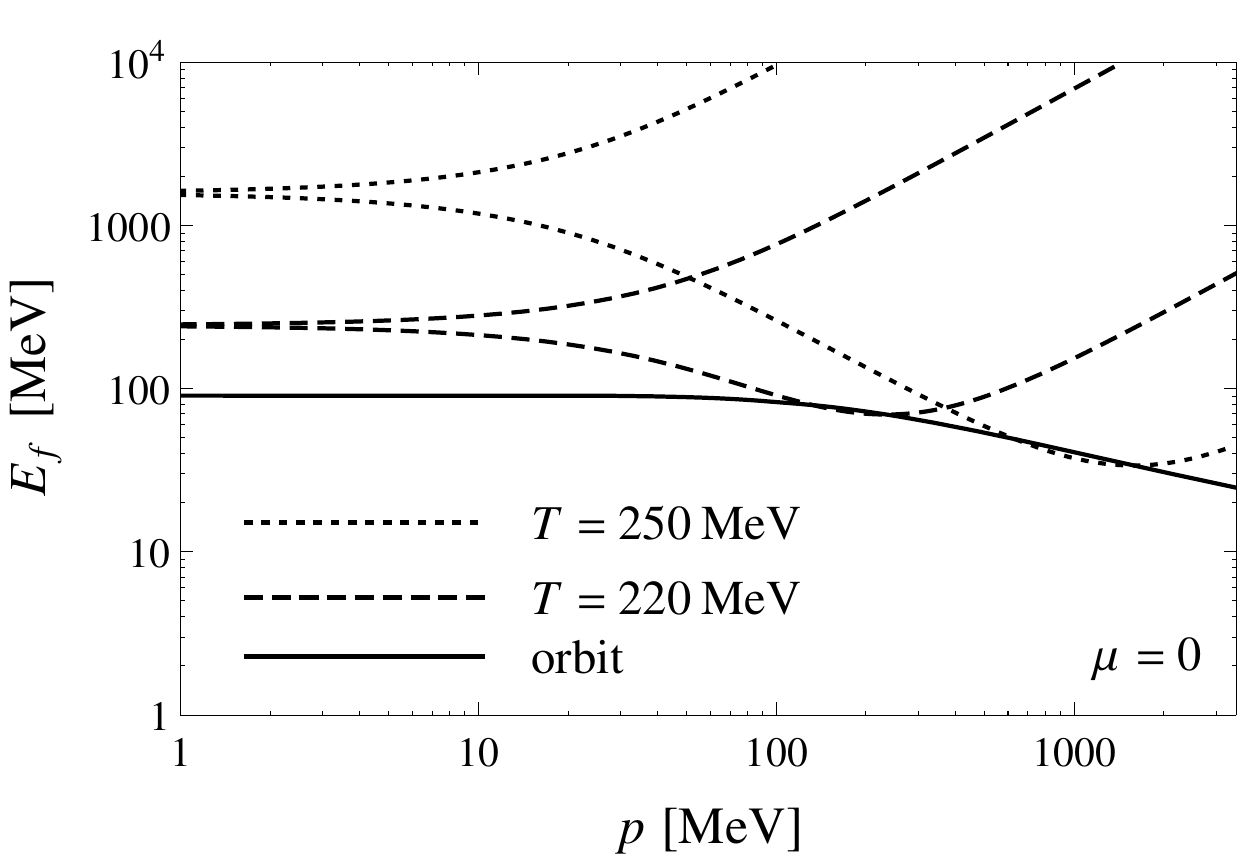}
  \caption{Range of integration for $E_f\in[E_{\rm min},E_{\rm max}]$ as function of absolute momentum $p=|\vec{p}|$ for different temperatures. The solid line displays the orbit of minimal values.}
  \label{Fig11}
\end{center}
\end{figure}
At given momentum $p$ the roots of $F(\,\cdot\, ,p)$ read, for the plus-sign case of both the $Z_1$ and $Z_2$ terms, where $E_b^2=(E_f+p_0)^2$:
\begin{equation}
\begin{aligned}
  E_{{\rm max},{\rm min}} &=\frac{1}{2M^2}\left[\left(m_{\rm M}^2-2M^2\right)\sqrt{M^2+p^2}\right. \\ 
   & \hspace{1.5cm} \left.\pm pm_{\rm M}\sqrt{m_{\rm M}^2-4M^2}\right].
\end{aligned}
\end{equation}
From this we find
\begin{equation}
\label{EmaxEminDifferenceZeroValue}
\begin{aligned}
  & E_{{\rm max}}-E_{{\rm min}}=\frac{p\,m_{\rm M}}{M^2}\sqrt{m_{\rm M}^2-4M^2}\sim p\;, \\
  & E_{{\rm max},{\rm min}}(p=0)=\frac{m_{\rm M}^2}{2M}-M>M\;.
\end{aligned}
\end{equation}
The momentum-dependent phase space for $E_f$ is shown in Fig.~\ref{Fig11} for different temperatures. Due to the larger meson mass at large $T$, the curves $E_{\rm min,max}$ are shifted to higher energies and momenta when increasing the temperature. There is an exact linear dependence of the integration range, $E_{\rm max}-E_{\rm min}$, on the incoming momentum $p$. At zero momentum the phase space collapses to one single point. Under the condition $m_{\rm M}>2M$ the phase space is always non-empty and compact with $M$ as minimal value of $E_{\rm min}$: $\emptyset\neq [E_{\rm min},E_{\rm max}]\subseteq [M,\infty)$. We also have derived the following substitution rule
\begin{equation}
  \int_{M}^\infty\frac{\dint E_f}{(2\pi)^3} \big(\cdot\big) \Theta(1-\xi^2) = \int_{E_{\rm min}}^{E_{\rm max}} \frac{\dint E_f}{(2\pi)^3} \big(\cdot\big)\;,
\end{equation}
which leads finally to a well-conditioned one-dimensional numerical integral.

Inserting the minus-sign case into Eq.\,\eqref{DefCapitalFfunctionPhaseSpace}, $E_b^2=(E_f-p_0)^2$, the phase space vanishes for any incoming momentum, since both the minimal and maximal energy of the loop fermion need to be negative:
\begin{equation}
  E_{\rm min}'=-E_{\rm max}\, ,\;\;\; E_{\rm max}'=-E_{\rm min}\;.
\end{equation}
We conclude that only the case $E_b=E_f+p_0$ leads to an on-shell contribution to the mesonic insertion. With this in mind the third term, involving $Z_3$, in Eq. \eqref{WidthFirstItThermalSelfEnergy} becomes:
\begin{equation}
\begin{aligned}
  \lim_{\eps\to 0} &\;\Im \left.\frac{\i\nu_n Z_3}{[(E_f+E_b)^2+\nu_n^2][(E_f-E_b)^2+\nu_n^2]}\right|_{\i\nu_n\mapsto p_0+\i\eps}= \\
   &=\frac{p_0\pi Z_3}{2p_0}\,\delta\big(\underbrace{[(E_f+E_b)^2-p_0^2]}_{=4E_fE_b}[E_b-E_f-p_0]\big)=\\
  &=\frac{\pi Z_3}{4E_f}\,\delta\left(E_b^2-(E_f+p_0)^2\right).
\end{aligned}
\end{equation}
Note that due to the $\i\nu_n$ factor in the first line, the $p_0$ terms cancel in the final result. As done in the calculation \eqref{CalcZ1Z2PhaseSpace} the momentum integral can be performed:
\begin{equation}
\begin{aligned}
  \int\frac{\dint^3q}{(2\pi)^3} \frac{\pi Z_3}{4E_f}\, &\delta(E_b^2-(E_f+p_0)^2)= \\
  &=2\pi\int_{M}^\infty\frac{\dint E_f}{(2\pi)^3}\frac{\pi Z_3}{8p}\,\Theta(1-\xi^2)\;.
\end{aligned}
\end{equation}
Combining all contributions, we find for the spectral width of an incoming quark:
\begin{equation}
\begin{aligned}
  &\Gamma^{\rm S/P}_{\rm q}(p) = \\
  &=-8\pi Mg_{\rm Mqq}^2\widetilde{N}^{\rm S/P}\!\int_{E_{\rm min}}^{E_{\rm max}}\frac{\dint E_f}{(2\pi)^3}\frac{\pi}{4|\vec{p}|}\left(\frac{E_f-E_b}{p_0}Z_2+\frac{Z_3}{2}\right)=\\
  &=\frac{Mg_{\rm Mqq}^2\widetilde{N}^{\rm S/P}}{4\pi|\vec{p}|}\int_{E_{\rm min}}^{E_{\rm max}}\dint E_f\left[n_{\rm B}(E_b)+n_{\rm F}^-(E_f)\right],
\end{aligned}
\end{equation}
using $E_b=E_f+p_0$. The effective spectral width in Eq.~\eqref{QuarkSpectralWidthOneLoopAllMesons} is determined by summing the pseudoscalar (pion) and scalar (sigma boson) channels weighted by their respective multiplicities, $3=N_{\rm f}^2-1$ and $1$.

\end{document}